\documentclass[12pt]{article}
\usepackage{amsfonts,euscript}
\begin{document}

\title{\begin{flushright}
{\normalsize Stockholm}\\[-0.3cm]{\normalsize USITP 03-10}\\[-0.3cm]
{\normalsize October 2003}\\[-0.3cm]
{\normalsize INSTITUT}\\[-0.3cm]
{\normalsize MITTAG-LEFFLER}
\end{flushright}
\vspace*{1cm}
Hamiltonian structure and noncommutativity in $p$-brane models with exotic supersymmetry}
\author{D.V.~Uvarov${}^a$ and A.A.~Zheltukhin${}^{a,b}$\\
{\normalsize ${}^a$ Kharkov Institute of Physics and Technology, 61108 Kharkov, Ukraine}\\
{\normalsize ${}^{b}$ Institute of Theoretical Physics, University of Stockholm, SCFAB,}\\
{\normalsize  SE-10691 Stockholm, Sweden}}
\date{}
\maketitle
\begin{abstract}
The Hamiltonian of the simplest super $p$-brane model preserving
$3/4$ of the $D=4$ $N=1$ supersymmetry in the centrally extended symplectic
 superspace is derived and its symmetries are described. The constraints of the model are covariantly separated into the first- and the second-class sets
and the Dirac brackets (D.B.) are constructed. We show the D.B. noncommutativity of the super $p$-brane coordinates and find the D.B.
 realization of the $OSp(1|8)$ superalgebra.
Established is the coincidence of the  D.B. and Poisson bracket realizations of the $OSp(1|8)$ superalgebra on the constraint surface and the absence there of anomaly terms in the commutation relations for the quantized generators of the superalgebra.
\end{abstract}

\section{Introduction}

As shown in \cite{ZU} exotic BPS states preserving $\frac{{\sf M}-1}{{\sf M}}$ fraction of $N=1$ supersymmetry can be realized by static configurations of free tensionless super $p$-branes $(p=1,2,...)$ with the action linear in derivatives\footnote{New Wess-Zumino like super $p$-brane models nonlinear in derivatives and preserving $\frac{{\sf M}-1}{{\sf M}}$ fraction of supersymmetry were recently proposed in \cite{BeZ}.}. These static configurations were described by  general solutions of the equations of motion of super $p$-branes evolving in superspace extended by tensor central charge (TCC) coordinates. Because of the $OSp(1|2{\sf M})$ global symmetry of the model, its static $p$-brane solution was formulated in terms of symplectic supertwistors previously used while studying superparticle models \cite{Fer}, \cite{Shir}, \cite{BL} and forming a subspace of the $Sp(2{\sf M})$ invariant symplectic space \cite{Fronsdal}, \cite{Vasiliev}. As a result, the static form of the discussed supertwistor representation of the BPS brane solution is not static in terms of the original superspace-time and TCC coordinates. It is static only modulo transformations of enhanced $\kappa$-symmetry and its accompanying local symmetries, since the supertwistor components are invariant under these gauge symmetries, as shown in \cite{ZU2}. The unphysical $p$-brane motions related to the gauge symmetries were geometrically realized as the abelian shifts \cite{ZU2} of the space-time and TCC coordinates by the Lorentz bivectors (generally multivectors) generalizing vector light-like Penrose shifts of the standard space-time coordinates \cite{Penrose}. Being inessential on the classical level of consideration, these shifts may turn out to be essential in the quantum dynamics of strings and branes. This necessitates quantum treatment of the model \cite{ZU} in the original variables that belong to the superspace extended by TCC coordinates and auxiliary spinor fields. An interest to this problem is stimulated by a conjectured relation of the tensionless strings with higher spin theories and free conformal SYM theories \cite{Bo}, \cite{Witten}, \cite{Vasiliev}, as well as by the presence of higher spins in the quantized $OSp(1|2{\sf M})$ invariant model of superparticle \cite{BLS}.

The Dirac analysis of the Hamiltonian structure of tensionless extended objects permits to outline some peculiarities of their quantum dynamics \cite{BZ90}, \cite{ILST}, \cite{ZL2}. For the case of dynamical systems including the second-class constraints, such  information is accumulated in the Dirac brackets used  during  the quantization procedure. The brane model \cite{ZU} contains the first- and the second-class constraints in view of the linear character of the Lagrangian in the world-volume time derivatives and the presence of the auxiliary spinor field parametrizing the string/brane momenta. Construction of the Dirac brackets  for the discussed super $p$-brane model implies its Hamiltonian analysis with a covariant division of the first- and the  second-class constraints.

As an example we solve this problem here for the $D=4$ $N=1$ super $p$-brane model  and construct its Hamiltonian and Lorentz covariant Dirac brackets. We find that the Hamiltonian symplectic structure of the brane  model encoded in the Dirac brackets is parametrized by only one  dynamical variable $\rho^{\tau}$ describing the proper time  component of the vector density $\rho^\mu$.
A covariant reduction  of the phase space excluding $\rho^\tau$ leads to the appearance of a nonlocal factor in the Dirac brackets depending on the
light-like projection of the super $p$-brane momentum.
It exposes an important distinction of the brane dynamics from the superparticle dynamics which may turn out to be essential in the  quantum picture.
We start the investigation of this problem and find  the Dirac bracket
(D.B.) noncommutativity between the space-time, TCC and auxiliary   brane coordinates.
To study effects of the noncommutativity on the algebraic level we construct the D.B. realization  of the $OSp(1|8)$ superalgebra of the global symmetry of the model. Established is the coincidence between the D.B. and P.B. realizations of the superalgebra, but only on the primary constraint surface. Applying the $\hat q\hat p$ ordering prescription,  previously studied in \cite{BZ90}, we consider a quantum realization of the $OSp(1|8)$ superalgebra and establish that its commutation relations are anomaly free on the constraint surface.

\section{Lagrangians for strings and branes with enhanced supersymmetry and
symplectic twistor}

A new simple model \cite{ZU} describes tensionless strings and $p-$branes spreading in the
symplectic superspace ${\cal M}^{susy}_{\sf M}$. For
${\sf M}=2^{[\frac{D}{2}]}$ $(D=2,3,4\ mod 8)$ this superspace naturally
associates with $D-$dimensional Minkowski space-time extended by the
Majorana spinor $\theta_a$ $(a=1,2,...,2^{[\frac{D}{2}]})$ and the tensor
central charge coordinates $z_{ab}$ additively unified with the standard
$x_{ab}=x^m(\gamma_mC^{-1})_{ab}$ space-time coordinates in the symmetric
spin-tensor $Y_{ab}$. The supersymmetric and reparametrization
invariant action of the model \cite{ZU}
\begin{equation}\label{1}
S_p=\frac12\int d\tau d^p\sigma\ \rho^\mu U^aW_{\mu ab}U^b
\end{equation}
includes the world-volume pullback
\begin{equation}\label{2}
W_{\mu ab}=\partial_\mu
Y_{ab}-2i(\partial_\mu\theta_a\theta_b+\partial_\mu\theta_b\theta_a)
\end{equation}
of the supersymmetric Cartan differential one-form $W_{ab}=W_{\mu
ab}d\xi^\mu$, where
$\partial_\mu\equiv\frac{\partial}{\partial\xi^\mu}$ and
$\xi^\mu=(\tau,\sigma^M)$, $M=1,2,...,p$ are world-volume
coordinates. The local auxiliary Majorana spinor
$U^a(\tau,\sigma^M)$ parametrizes the generalized momentum
$P^{ab}=\rho^{\tau}U^aU^b$ of the tensionless $p-$brane and
$\rho^\mu(\tau,\sigma^M)$ is the world-volume vector  density providing
the reparametrization invariance of $S_p$. This action has $({\sf
M}-1)$ $\kappa-$symmetries and consequently preserves $\frac{{\sf
M}-1}{{\sf M}}$ fraction of the original global supersymmetry.

By the generalized Penrose transformation of variables
\begin{equation}\label{3}
Y_{ab}U^b=i\tilde
Y_a+\tilde\eta\theta_a,\quad\tilde\eta=-2i(U^a\theta_a),
\end{equation}
where $\tilde\eta$ is real Goldstone fermion associated with the spontaneous
breakdown of $\frac{1}{{\sf M}}$ supersymmetry, the differential one-form
$U^aW_{ab}U^b$ is presented as
\begin{equation}\label{4}
U^aW_{ab}U^b=i\{U^ad\tilde Y_a-dU^a\tilde Y_a+d\tilde\eta\tilde\eta\}\equiv
dY^\Lambda G_{\Lambda\Sigma}Y^{\Sigma}.
\end{equation}
The new object $Y^\Lambda=(iU^a, \tilde Y_a, \tilde\eta)$ in (\ref{3}), (\ref{4}) is
$OSp(1|2{\sf M})$ supertwistor and
$G_{\Lambda\Sigma}=(-)^{\Lambda\Sigma+1}G_{\Sigma\Lambda}$ is $OSp(1|2{\sf
M})$ invariant supersymplectic metric
\begin{equation}\label{5}
G_{\Lambda\Sigma}=\left(\begin{array}{cc} \omega^{(2{\sf M})}&0\\0&i
\end{array}\right)=\left(\begin{array}{ccc} 0&-\delta_a{}^b&0\\
\delta^a{}_b&0&0\\ 0&0&i\end{array}\right),
\end{equation}
which is the supersymmetric generalization of $Sp(2{\sf M})$ symplectic metric
$\omega^{(2{\sf M})}$. In view of (\ref{3}) and (\ref{4}), the action $S_p$
(\ref{1}) is presented in the supertwistor form
\begin{equation}\label{6}
S_p=\frac12\int d\tau d^p\sigma\ \rho^\mu\partial_\mu Y^\Lambda
G_{\Lambda\Sigma}Y^\Sigma
\end{equation}
that is apparently invariant under the global generalized
superconformal $OSp(1|2{\sf M})$ symmetry. For the particular case
of $D=11$ the action (\ref{6}) invariant under the $OSp(1|64)$
symmetry  was considered in \cite{Bandos}.

The original action (\ref{1}) is invariant under $({\sf M}-1)$
$\kappa-$symmetries since the transformation parameter $\kappa_a$ is restricted by only one real
condition
\begin{equation}\label{7}
U^a\kappa_a=0,
\end{equation}
as it follows from the transformation rules of the primary variables
\begin{equation}\label{8}
\delta_\kappa\theta_a=\kappa_a,\quad\delta_\kappa
Y_{ab}=-2i(\theta_a\kappa_b+\theta_b\kappa_a),\quad\delta_\kappa
U^a=0.
\end{equation}\label{8'}
It is easy to show that all components of the supertwistor
$Y^\Lambda=(iU^a, \tilde Y_a, \tilde\eta)$ are invariant under
$\kappa-$symmetry transformations (\ref{7}), (\ref{8})
\begin{equation}
\delta_\kappa\tilde Y_a=0,\quad\delta_\kappa\tilde\eta=0,\quad\delta_\kappa
U^a=0,
\end{equation}
so that the new representation of $S_p$ (\ref{6}) includes only $\kappa-$invariant
variables.  Note that in $4-$dimensional space-time $Y^\Lambda$
contains only 9 real variables that is twice less than the number of the
original variables $Y_{ab}$, $\theta_a$, $U^a$.

\section{Example of the  $OSp(1|8)$ invariant string/brane model.
Primary constraints}

$OSp(1|8)$ is the global supersymmetry
of the massless fields of all spins in $D=4$ space-time extended by TCC
coordinates \cite{Fronsdal}, \cite{Vasiliev}. Therefore, we study $D=4$
example of the string/brane model (\ref{1}) formulated in generalized $(4+6)-$dimensional
space ${\cal M}_{4+6}$ extended by the Grassmannian Majorana bispinor $\theta_a$. In
this case the $D=4$ $N=1$ superalgebra
\begin{equation}\label{9}
\{Q_a,Q_b\}=(\gamma^m C^{-1})_{ab}P_m+i(\gamma^{mn}C^{-1})_{ab}Z_{mn}
\end{equation}
includes the TCC two-form $Z_{mn}$, and the matrix coordinates $Y_{ab}$ are
\begin{equation}
Y_{ab}=x_{ab}+z_{ab},
\end{equation}
where
\begin{equation}\label{10}
x_{ab}=x_m(\gamma^m C^{-1})_{ab},\quad z_{ab}=z_{mn}(\gamma^{mn}C^{-1})_{ab}
\end{equation}
with the charge conjugation matrix $C$ chosen to be imaginary in the
Majorana representation. Here we use the same agreements about the spinor
algebra as in \cite{ZU}.

In the Weyl basis real symmetric $4\times4$ matrix
$Y_{ab}$ is presented as
\begin{equation}\label{11}
Y_a{}^b=Y_{ad}C^{db}=\left(\begin{array}{cc} z_\alpha{}^\beta&
x_{\alpha\dot\beta}\\ \tilde x^{\dot\alpha\beta}& \bar
z^{\dot\alpha}{}_{\dot\beta}\\ \end{array}\right),
\end{equation}
\begin{equation}\label{14'}
C^{ab}=\left(\begin{array}{cc} \epsilon^{\alpha\beta}& 0\\ 0&
\epsilon_{\dot\alpha\dot\beta} \end{array}\right).
\end{equation}
In the $D=4$ case the auxiliary Majorana spinor
$U^a(\tau,\sigma^M)$ together with
other auxiliary Majorana spinors $V^a(\tau,\sigma^M)$,
 $(\gamma_5U)_a$ and $(\gamma_5V)_a$ form a local spinor basis
\begin{equation}\label{15}
U_a={u_\alpha\choose \bar u^{\dot\alpha}},\quad V_a={v_\alpha\choose \bar
v^{\dot\alpha}},\quad(U\gamma_5V)=-2i,\quad (UV)=0
\end{equation}
attached to string/brane world volume and the $\gamma_5-$matrix is
\begin{equation}
(\gamma_{5})_a{}^b=\left(\begin{array}{cc} -i\delta_\alpha^\beta&0\\
0&i\delta^{\dot\alpha}_{\dot\beta} \end{array}\right).
\end{equation}
Respectively the linear independent Weyl spinors $u^\alpha$ and $v^\alpha$
 may be identified with the local Neuman-Penrose dyad \cite{Penrose}
\begin{equation}\label{17}
u^\alpha v_\alpha\equiv u^\alpha\varepsilon_{\alpha\beta}v^\beta=1,\quad
u^\alpha u_\alpha=v^\alpha v_\alpha=0.
\end{equation}

In the Weyl basis the action (\ref{1}) acquires the form
\begin{equation}\label{14}
S_p=\frac12\int d\tau d^p\sigma\ \rho^\mu
\left(2u^\alpha\omega_{\mu\alpha\dot\alpha}\bar
u^{\dot\alpha}+u^\alpha\omega_{\mu\alpha\beta}u^\beta+\bar
u^{\dot\alpha}\bar\omega_{\mu\dot\alpha\dot\beta}\bar u^{\dot\beta}\right),
\end{equation}
where the supersymmetric one-forms $\omega_{\mu\alpha\dot\alpha}$ and
$\omega_{\mu\alpha\beta}$ are
\begin{equation}\begin{array}{c}
\omega_{\mu\alpha\dot\alpha}=\partial_\mu
x_{\alpha\dot\alpha}+2i(\partial_\mu\theta_\alpha\bar\theta_{\dot\alpha}+\partial_\mu\bar\theta_{\dot\alpha}\theta_\alpha),\\[0.2cm]
\omega_{\mu\alpha\beta}=-\partial_\mu
z_{\alpha\beta}-2i(\partial_\mu\theta_\alpha\theta_\beta+\partial_\mu\theta_\beta\theta_\alpha),\\[0.2cm]
\bar\omega_{\mu\dot\alpha\dot\beta}=-\partial_\mu\bar
z_{\dot\alpha\dot\beta}-2i(\partial_\mu\bar\theta_{\dot\alpha}\bar\theta_{\dot\beta}+\partial_\mu\bar\theta_{\dot\beta}\bar\theta_{\dot\alpha}).\\
\end{array}
\end{equation}
The momenta densities ${\EuScript P}^{\mathfrak M}(\tau,\sigma^M)$
\begin{equation}\label{20}
{\EuScript P}^{\mathfrak M}=\frac{\partial L}{\partial\dot\EuScript
Q_{\mathfrak M}}=(P^{\dot\alpha\alpha}, \pi^{\alpha\beta},
\bar\pi^{\dot\alpha\dot\beta}, \pi^\alpha, \bar\pi^{\dot\alpha}, P^\alpha_u,
\bar P^{\dot\alpha}_u, P^\alpha_v, \bar P^{\dot\alpha}_v, P^{(\rho)}_\mu)
\end{equation}
are canonically conjugate to the coordinates
\begin{equation}\label{21}
\EuScript Q_{\mathfrak M}=(x_{\alpha\dot\alpha}, z_{\alpha\beta}, \bar
z_{\dot\alpha\dot\beta}, u_\alpha, \bar u_{\dot\alpha}, v_\alpha, \bar
v_{\dot\alpha}, \rho^{\mu})
\end{equation}
with respect to the Poisson brackets
\begin{equation}
\{{\EuScript P}^{\mathfrak M}(\vec\sigma),\EuScript Q_{\mathfrak
N}(\vec\sigma')\}_{P.B.}=\delta^{\mathfrak M}_{\mathfrak
N}\delta^p(\vec\sigma-\vec\sigma ')
\end{equation}
with the periodic $\delta-$function $\delta^p(\vec\sigma-\vec\sigma')$, where $\vec\sigma=(\sigma^1,...,\sigma^p)$, for
the case of closed string/brane studied here.

As far as $S_p$ (\ref{14}) is linear in the proper time derivatives, it is
characterized by the presence of the primary constraints. These constraints
may be divided into four sectors.

The bosonic $\Phi-$sector includes the
constraints $\Phi\equiv(\Phi^{\dot\alpha\alpha}, \Phi^{\alpha\beta},
\bar\Phi^{\dot\alpha\dot\beta})$ with
\begin{equation}\label{23}\begin{array}{c}
\Phi^{\dot\alpha\alpha}=P^{\dot\alpha\alpha}-\rho^\tau u^\alpha
\bar u^{\dot\alpha}\approx0,\\[0.2cm]
\Phi^{\alpha\beta}=\pi^{\alpha\beta}+\frac12\rho^{\tau}u^\alpha
u^\beta\approx0,\\[0.2cm]
\bar\Phi^{\dot\alpha\dot\beta}=\bar\pi^{\dot\alpha\dot\beta}+\frac12\rho^{\tau}\bar u^{\dot\alpha}
\bar u^{\dot\beta}\approx 0.
\end{array}\end{equation}
The constraints from the Grassmannian $\Psi-$sector, where
$\Psi=(\Psi^\alpha, \bar\Psi^{\dot\alpha})$, are given by
\begin{equation}\label{24}\begin{array}{c}
\Psi^\alpha=\pi^\alpha-2i\bar\theta_{\dot\alpha}P^{\dot\alpha\alpha}-4i\pi^{\alpha\beta}\theta_\beta\approx0,\\
\bar\Psi^{\dot\alpha}=-(\Psi^\alpha)^\ast=\bar\pi^{\dot\alpha}-2iP^{\dot\alpha\alpha}\theta_\alpha-4i\bar\pi^{\dot\alpha\dot\beta}\bar\theta_{\dot\beta}\approx0.
\end{array}
\end{equation}
The dyad or $(u,v)-$sector is formed by the constraints
\begin{equation}\label{25}\begin{array}{c}
P^\alpha_u\approx0,\quad\bar P^{\dot\alpha}_u\approx0,\quad
P^\alpha_v\approx0,\quad\bar P^{\dot\alpha}_v\approx0,\\
\Xi\equiv u^\alpha v_\alpha-1\approx0,\quad\bar\Xi\equiv\bar
u^{\dot\alpha}\bar v_{\dot\alpha}-1\approx0.
\end{array}
\end{equation}
Finally, the $\rho-$sector includes the constraints
\begin{equation}\label{26}
P^{(\rho)}_\mu\approx0,\quad\mu=(\tau,M),\quad M=(1,...,p).
\end{equation}
The constraints forming $\Phi-$sector have zero Poisson
brackets  (P.B.) among themselves and with the $\Psi-$sector constraints
\begin{equation}\label{27}
\{\Phi(\vec\sigma),\Phi(\vec\sigma
')\}_{P.B.}=0,\quad\{\Phi(\vec\sigma),\Psi(\vec\sigma ')\}_{P.B.}=0.
\end{equation}
They also P.B. commute with $P^\alpha_v$, $\bar P^{\dot\alpha}_v$, $\Xi$,
$\bar\Xi$ and $P^{(\rho)}_M$
\begin{equation}\label{28}
\{\Phi(\vec\sigma),P^\alpha_v(\vec\sigma')\}_{P.B.}=0,\quad\{\Phi(\vec\sigma),\Xi(\vec\sigma')\}_{P.B.}=0,\quad\{\Phi(\vec\sigma),P^{(\rho)}_{M}(\vec\sigma')\}_{P.B.}=0.
\end{equation}

The $\Psi-$constraints have  zero P.B. with the constraints from all other sectors, but have nonzero P.B. among themselves.
Note that the constraints (\ref{23})-(\ref{26}) do not contain the
space-time $x_{\alpha\dot\alpha}$ and TCC $z_{\alpha\beta}, \bar z_{\dot\alpha\dot\beta}$ coordinates  as well as the $\rho^{M}$ components.

To find all local symmetries of the brane action, it is necessary
to split the constraints (\ref{23})-(\ref{26}) into the first- and
second-class constraints. Then the P.B. of the first-class
constraints with the brane coordinates will generate the local
symmetries in accordance with the Dirac prescription.

\section{$\Psi-$sector: $(3\oplus1)-$splitting and first-class
constraints generating enhanced $\kappa-$symmetry}

Here we find the Hamiltonian realization of the enhanced $\kappa-$symmetry
generators. The generators of $\kappa-$symmetry are included into the
$\Psi-$sector (\ref{24}) of the primary constraints due to their Grassmannian
graduation. To derive these constraints it is enough to study only Poisson
brackets among the $\Psi-$constraints
\begin{equation}\label{30}
\{\Psi^\alpha(\vec\sigma),\Psi^\beta(\vec\sigma{}')\}_{P.B.}=-8i\pi^{\alpha\beta}\delta^p(\vec\sigma-\vec\sigma{}')=-8i\left(\Phi^{\alpha\beta}-\frac12\rho^\tau
u^\alpha u^\beta\right)\delta^p(\vec\sigma-\vec\sigma{}'),
\end{equation}
\begin{equation}\label{31}
\{\Psi^\alpha(\vec\sigma),\bar\Psi^{\dot\beta}(\vec\sigma{}')\}_{P.B.}=-4iP^{\dot\beta\alpha}\delta^p(\vec\sigma-\vec\sigma{}')=-4i\left(\Phi^{\dot\beta\alpha}+\rho^\tau
u^\alpha\bar u^{\dot\beta}\right)\delta^p(\vec\sigma-\vec\sigma{}'),
\end{equation}
because of the commutativity of the $\Psi-$sector with the others. Then upon
the multiplication of (\ref{30}), (\ref{31}) by $u_\beta(\tau,\vec\sigma')$ and
$\bar u_{\dot\beta}(\tau,\vec\sigma')$, respectively, and using (\ref{17}) we
find
\begin{equation}
\{\Psi^\alpha(\vec\sigma),\Psi^{(u)}(\vec\sigma')\}_{P.B.}=-8i\Phi^{\alpha\beta}u_\beta\delta^p(\vec\sigma-\vec\sigma'),\quad
\{\Psi^\alpha(\vec\sigma),\bar\Psi^{(u)}(\vec\sigma')\}_{P.B.}=-4i\bar u_{\dot\beta}\Phi^{\dot\beta\alpha}\delta^p(\vec\sigma-\vec\sigma').
\end{equation}
This means that $\Psi^{(u)}$ and $\bar\Psi^{(u)}$
\begin{equation}
\Psi^{(u)}\equiv\Psi^\alpha
u_\alpha\approx0,\quad\bar\Psi^{(u)}\equiv\bar\Psi^{\dot\alpha}
\bar u_{\dot\alpha}\approx0
\end{equation}
are the first-class constraints. The Poisson brackets between the
$\Psi^{(u)}$, $\bar\Psi^{(u)}$ are the following
\begin{equation}\label{34}
\begin{array}{c}
\{\Psi^{(u)}(\vec\sigma),\bar\Psi^{(u)}(\vec\sigma')\}_{P.B.}=-4i\Phi^{(u)}\delta^p(\vec\sigma-\vec\sigma')\approx0,\\[0.2cm]
\{\Psi^{(u)}(\vec\sigma),\Psi^{(u)}(\vec\sigma')\}_{P.B.}=-8iT^{(u)}\delta^p(\vec\sigma-\vec\sigma')\approx0,\\[0.2cm]
\{\bar\Psi^{(u)}(\vec\sigma),\bar\Psi^{(u)}(\vec\sigma')\}_{P.B.}=-8i\bar
T^{(u)}\delta^p(\vec\sigma-\vec\sigma')\approx0,\\
\end{array}
\end{equation}
where $T^{(u)}$ and $\Phi^{(u)}$ are the constraints
\begin{equation}\label{35}
T^{(u)}\equiv u_\alpha\Phi^{\alpha\beta}u_\beta\approx0,\quad\bar
T^{(u)}=(T^{(u)})^\ast,\quad
\Phi^{(u)}\equiv
\bar u_{\dot\beta}
\Phi^{\dot\beta\alpha}u_\alpha.
\end{equation}

The remaining two constraints from the $\Psi-$sector are presented by the
projections
\begin{equation}
\Psi^{(v)}\equiv \Psi^\alpha v_\alpha,\quad\bar\Psi^{(v)}\equiv
\bar\Psi^{\dot\alpha}\bar v_{\dot\alpha}.
\end{equation}
Projecting the Poisson brackets (\ref{30}), (\ref{31}) on the spinors
$v_\beta(\tau,\vec\sigma')$ and $\bar v_{\dot\beta}(\tau,\vec\sigma')$ and summing up the resulting expressions
we find
\begin{equation}\label{34'}
\{\Psi^\alpha(\vec\sigma),\Psi^{(v)}_R(\vec\sigma')\}_{P.B.}=-4i(\bar v_{\dot\beta}\Phi^{\dot\beta\alpha}
+2\Phi^{\alpha\beta}v_\beta)\delta^p(\vec\sigma-\vec\sigma')\approx0,
\end{equation}
where the real constraint $\Psi^{(v)}_R$ is defined by the sum
\begin{equation}
\Psi^{(v)}_R\equiv\Psi^{(v)}+\bar\Psi^{(v)}\approx0.
\end{equation}
The Poisson brackets (\ref{34'}) show that $\Psi^{(v)}_R$ is the real
first-class constraint. 
 
Multiplying (\ref{34'}) by $u_\alpha(\tau,\vec\sigma)$ we find
\begin{equation}
\begin{array}{c}
\{\Psi^{(u)}(\vec\sigma), \Psi^{(v)}_R(\vec\sigma')\}_{P.B.}=-4i(\bar v_{\dot\beta}\Phi^{\dot\beta\alpha}u_\alpha
+2u_\alpha\Phi^{\alpha\beta}v_{\beta})\delta^p(\vec\sigma-\vec\sigma')\approx0,\\[0.2cm]
\{\bar\Psi^{(u)}(\vec\sigma), \Psi^{(v)}_R(\vec\sigma')\}_{P.B.}=-4i(\bar u_{\dot\beta}\Phi^{\dot\beta\alpha}v_\alpha
+2\bar u_{\dot\alpha}\bar\Phi^{\dot\alpha\dot\beta}\bar
v_{\dot\beta})\delta^p(\vec\sigma-\vec\sigma')\approx0.
\end{array}
\end{equation}
The Poisson brackets for $\Psi^{(v)}_R$ with itself are
\begin{equation}\label{37}
\{\Psi^{(v)}_R(\vec\sigma),\Psi^{(v)}_R(\vec\sigma')\}_{P.B.}=-8i\tilde
T^{(v)}_R\delta^p(\vec\sigma-\vec\sigma')\approx0,
\end{equation}
where the real constraint $\tilde T^{(v)}_R$ from the $\Phi-$sector is
defined as
\begin{equation}\label{38}
\begin{array}{c}
\tilde T^{(v)}_R\equiv T^{(v)}_R+\Phi^{(v)}=T^{(v)}+\bar
T^{(v)}+\Phi^{(v)}\approx0,\\
T^{(v)}\equiv v_\alpha \Phi ^{\alpha\beta}v_\beta, \quad
\Phi^{(v)}\equiv \bar v_{\dot\alpha}\Phi^{\dot\alpha\alpha}v_\alpha.
\end{array}
\end{equation}
 Thus, the $\Psi-$sector
constraints are split into the three real constraints $\Psi^{(u)}$,
$\bar\Psi^{(u)}$ and $\Psi^{(v)}_R$ belonging to the first-class and one
real constraint $\Psi^{(v)}_I$ given by the imaginary part of $\Psi^{(v)}$
\begin{equation}\label{38'}
\Psi^{(v)}_I=i(\Psi^\alpha v_\alpha-\bar\Psi^{\dot\alpha}\bar
v_{\dot\alpha})\approx0.
\end{equation}
The calculation of the Poisson brackets for $\Psi^{(v)}_I$ with itself
yields
\begin{equation}
\{\Psi^{(v)}_I(\vec\sigma),\Psi^{(v)}_I(\vec\sigma')\}_{P.B.}=8i(T^{(v)}_R-\Phi^{(v)}-2\rho^\tau)\delta^p(\vec\sigma-\vec\sigma'),
\end{equation}
resulting to the weak equality
\begin{equation}
\{\Psi^{(v)}_I(\vec\sigma),\Psi^{(v)}_I(\vec\sigma')\}_{P.B.}\approx-16i\rho^\tau\delta^p(\vec\sigma-\vec\sigma'),
\end{equation}
proving that $\Psi^{(v)}_I$ is the second-class constraint.
Therefore, value of the nonzero world-volume field $\rho^\tau$ measures breaking of the fourth $\kappa$-symmetry of the brane model.

Using the definition of canonical Poisson brackets
\begin{equation}
\{\pi^\alpha(\vec\sigma),\theta_\beta(\vec\sigma')\}_{P.B.}=\delta^\alpha_\beta\delta^p(\vec\sigma-\vec\sigma'),\quad\{\bar\pi^{\dot\alpha}(\vec\sigma),\bar\theta_{\dot\beta}(\vec\sigma')\}_{P.B.}=\delta^{\dot\alpha}_{\dot\beta}\delta^p(\vec\sigma-\vec\sigma')
\end{equation}
we find the transformations of the $\theta-$coordinates under the charges
corresponding to the first-class constraints $\Psi^{(u)}$, $\bar\Psi^{(u)}$
and $\Psi^{(v)}_R$
\begin{equation}\label{43}
\begin{array}{c}
\delta_\kappa\theta_\alpha=\{\int
d^p\sigma'\kappa\Psi^{(u)}(\vec\sigma'),\theta_\alpha(\vec\sigma)\}_{P.B.}=\kappa
u_\alpha,\\[0.2cm]
\delta_\kappa\bar\theta_{\dot\alpha}=\{\int
d^p\sigma'\bar\kappa\bar\Psi^{(u)}(\vec\sigma'),\bar\theta_{\dot\alpha}(\vec\sigma)\}_{P.B.}=\bar\kappa
\bar u_{\dot\alpha},\\[0.2cm]
\delta_\kappa x_{\alpha\dot\alpha}=-2i(\kappa
u_\alpha\bar\theta_{\dot\alpha}
+\bar\kappa\bar u_{\dot\alpha}\theta_{\alpha})
,\quad
\delta_\kappa z_{\alpha\beta}=-2i\kappa(
u_\alpha\theta_{\beta}
+u_\beta\theta_{\alpha})
;\\[0.2cm]
\delta_{\kappa_R}\theta_\alpha=\{\int
d^p\sigma'\kappa_R\Psi^{(v)}_R(\vec\sigma'),\theta_\alpha(\vec\sigma)\}_{P.B.}=\kappa_R
v_\alpha,\\[0.2cm]
\delta_{\kappa_R}\bar\theta_{\dot\alpha}=\{\int
d^p\sigma'\kappa_R\Psi^{(v)}_R(\vec\sigma'),\bar\theta_{\dot\alpha}(\vec\sigma)\}_{P.B.}=\kappa_R
\bar v_{\dot\alpha},\\[0.2cm]
\delta_{\kappa_R} x_{\alpha\dot\alpha}=-2i\kappa_R(
v_\alpha\bar\theta_{\dot\alpha}+\bar v_{\dot\alpha}\theta_\alpha),\quad\delta_{\kappa_R}
z_{\alpha\beta}=-2i\kappa_R( v_\alpha\theta_{\beta}+v_\beta\theta_{\alpha}).
\end{array}
\end{equation}
To connect the transformations (\ref{43}) with the original
$\kappa-$symmetry transformations (\ref{8}) let us expand $\kappa_\alpha$ in the
dyad basis
\begin{equation}
\kappa_\alpha=-(\kappa_\beta v^\beta)u_\alpha+(\kappa_\beta
u^\beta)v_\alpha=-(\kappa_\beta v^\beta)u_\alpha+[Re(\kappa_\beta
u^\beta)+iIm(\kappa_\beta u^\beta)]v_\alpha.
\end{equation}
Then the transformations (\ref{8}) are presented in the form
\begin{equation}\label{45}
\delta_{\kappa_\alpha}\theta_\alpha=-(\kappa_\beta
v^\beta)u_\alpha+Re(\kappa_\beta u^\beta)v_\alpha,
\end{equation}
in view of the condition $Im(\kappa_\beta u^\beta)=0$ equivalent to (\ref{7}).  Comparison of
(\ref{43}) and (\ref{45}) shows that the transformations (\ref{43}) are the
original $\kappa-$symmetry transformations with the complex parameter
$\kappa$ and the real parameter $\kappa_R$ connected with the original
parameters $\kappa_\alpha$ by the relations
\begin{equation}
\kappa=-\kappa_\alpha v^\alpha,\quad\kappa_R=Re(\kappa_\alpha u^\alpha).
\end{equation}
Therefore,  we proved that the first-class constraints $\Psi^{(u)}$,
$\bar\Psi^{(u)}$ and $\Psi^{(v)}_R$ are the generators of three
$\kappa-$symmetries since their Poisson brackets are closed by the
constraints from the $\Phi-$sector. We shall comment these Poisson brackets
in the next section, where the division of the $\Phi-$sector into the first-
and the second-class constraints will be considered.

\section{$\Phi-$sector: $(6\oplus4)-$splitting and first-class
constraints generating new local symmetries}

The Poisson brackets (\ref{34}) of the $\kappa-$symmetry constraints $\Psi^{(u)}$
and $\bar\Psi^{(u)}$ are closed by the constraints $\Phi^{(u)}$ and
$T^{(u)}$ (\ref{35}) from the $\Phi-$sector. Let us show that the latter
constraints are also the first-class ones. It is easy to see that the
constraint $\Phi^{(u)}$ transforms the $x-$coordinates but leaves  all
other variables in $S_p$ (\ref{14}) intact. The transformation of the $x-$coordinates
is
\begin{equation}\label{47}
\delta_{\Phi^{(u)}} x_{\alpha\dot\alpha}=\{\int
d^p\sigma'\epsilon_{\Phi^{(u)}}\Phi^{(u)}(\vec\sigma'),
x_{\alpha\dot\alpha}(\vec\sigma)\}_{P.B.}=\epsilon_{\Phi^{(u)}}u_\alpha\bar
u_{\dot\alpha}
\end{equation}
and is the local symmetry of the action $S_p$ (\ref{14})
\begin{equation}
\delta_{\Phi^{(u)}}S_p=\int d\tau d^p\sigma\rho^\mu
u^\alpha\partial_\mu(\epsilon_{\Phi^{(u)}}u_\alpha\bar u_{\dot\alpha})\bar
u^{\dot\alpha}=0
\end{equation}
due to the relation (\ref{17}) $u^\alpha u_\alpha=0$. Consequently, the
constraint $\Phi^{(u)}$ is the first-class one. The transformation
(\ref{47}) is the local shift of $x_m$ along the light-like $4-$vector
$(u\sigma_m\bar u)$
\begin{equation}\label{48}
\delta_{\Phi^{(u)}}
x_{m}=-\frac12\epsilon_{\Phi^{(u)}}(u\sigma_m\bar u)
\end{equation}
and can be rewritten as a weak equality
\begin{equation}\label{49}
\delta_{\Phi^{(u)}}
x_{m}\approx\tilde\epsilon_{\Phi^{(u)}}P_m,
\end{equation}
where $P_m=-2P^{\dot\alpha\alpha}\sigma_{m\alpha\dot\alpha}$ and
$\tilde\epsilon_{\Phi^{(u)}}\equiv\frac{\epsilon_{\Phi^{(u)}}}{\rho^\tau}$.
On the contrary, the constraints $T^{(u)}$ and $\bar T^{(u)}$ change only
the TCC coordinates $z_\alpha{}^\beta$
\begin{equation}\label{50}
\delta_{T^{(u)}} z_{\alpha\beta}=\epsilon_{T^{(u)}} u_\alpha u_\beta,\quad
\delta_{\bar T^{(u)}}\bar z_{\dot\alpha\dot\beta}=\bar\epsilon_{\bar
T^{(u)}}\bar u_{\dot\alpha} \bar u_{\dot\beta},
\end{equation}
as it follows after utilization of the canonical Poisson brackets
\begin{equation}
\{\pi^{\alpha\beta}(\vec\sigma),z_{\gamma\delta}(\vec\sigma')\}_{P.B.}=\frac12(\delta^\alpha_\gamma\delta^\beta_\delta+\delta^\alpha_\delta\delta^\beta_\gamma)\delta^p(\vec\sigma-\vec\sigma').
\end{equation}
The transformations (\ref{50}) do not change the action (\ref{14})
\begin{equation}
\delta_{T^{(u)}} S_p=\frac12\int d\tau d^p\sigma\rho^\mu[u^\alpha\partial_\mu(\epsilon
_{T^{(u)}}u_\alpha u_\beta)u^\beta+c.c.]=0
\end{equation}
and, consequently, are the first-class constraints too.Ê\\
To establish the
geometric sense of the transformations (\ref{50}) let us multiply them by
$(\sigma_{mn}\varepsilon)^{\alpha\beta}$ and
$(\tilde\sigma_{mn}\varepsilon)^{\dot\alpha\dot\beta}$, respectively, and sum
up the results. Then we find
\begin{equation}\label{53}
\delta_{T^{(u)}} z_{mn}=-\frac{i}{4}[\epsilon^{(R)}_{T^{(u)}}(u\sigma_{mn}u+\bar
u\tilde\sigma_{mn}\bar u)+\epsilon^{(I)}_{T^{(u)}}(u\sigma_{mn}u-\bar
u\tilde\sigma_{mn}\bar u)]
\end{equation}
 using the relation \cite{zli2}
\begin{equation}
 z_{mn}=-\frac{i}{4}[z_\alpha{}^\beta\sigma_{mn\beta}{}^\alpha+\bar
z^{\dot\beta}{}_{\dot\alpha}\tilde\sigma_{mn}{}^{\dot\alpha}{}_{\dot\beta}]
\end{equation}
connecting the spinor representation $z_{\alpha\beta}$ of the TCC
coordinates with the tensor representation) $z_{mn}=-z_{nm}$. In terms of the
Majorana spinor $U_a$ (\ref{15}) the transformation (\ref{53}) is presented
as
\begin{equation}
\delta_{T^{(u)}} z_{mn}=\frac{i}{8}[\epsilon^{(R)}_{T^{(u)}}(\bar
U\gamma_{mn}U)+\epsilon^{(I)}_{T^{(u)}}(\bar U\gamma_{mn}\gamma_5 U)],
\end{equation}
with the real parameters $\epsilon^{(R)}_{T^{(u)}}$ and $\epsilon^{(I)}_{T^{(u)}}$
\begin{equation}
\epsilon^{(R)}_{T^{(u)}}=\frac12(\epsilon_{T^{(u)}}+\bar\epsilon_{\bar
T^{(u)}}),\quad\epsilon^{(I)}_T=\frac{1}{2i}(\epsilon_{T^{(u)}}-\bar\epsilon_{\bar
T^{(u)}}).
\end{equation}
Now let us take into account the observation \cite{GZ} that the bivectors
$(\bar U\gamma_{mn}U)$ and $(\bar U\gamma_{mn}\gamma_5U)$ are null tensors,
i.e.
\begin{equation}\label{57}
(\bar U\gamma_{mn}U)^2=0,\quad(\bar U\gamma_{mn}\gamma_5U)^2=0.
\end{equation}
This means that $T^{(u)}$ and $\bar T^{(u)}$ generate the local shifts of
$z_{mn}$ along the isotropic bivectors (\ref{57}) and these shifts are a
natural generalization of the vector light-like shift (\ref{48}). On the
other hand, these shifts may be presented as the local shifts along the TCC
momentum
\begin{equation}
\delta_{T^{(u)}}z_{\alpha\beta}\approx-2\tilde\epsilon_{T^{(u)}}\pi_{\alpha\beta},\quad\delta_{\bar
T^{(u)}}\bar
z_{\dot\alpha\dot\beta}\approx-2\bar{\tilde\epsilon}_{\bar T^{(u)}}\bar\pi_{\dot\alpha\dot\beta}
\end{equation}
if the primary constraints (\ref{23}) are taken into account. Thus, the
Poisson brackets (\ref{34}) can be presented in the form including only the
first-class constraints
\begin{equation}\label{59}
\begin{array}{c}
\{\Psi^{(u)}(\vec\sigma),\bar\Psi^{(u)}(\vec\sigma')\}_{P.B.}=-8i(uP\bar
u)\delta^p(\vec\sigma-\vec\sigma'),\\[0.2cm]
\{\Psi^{(u)}(\vec\sigma),\Psi^{(u)}(\vec\sigma')\}_{P.B.}=-4i(u\pi
u)\delta^p(\vec\sigma-\vec\sigma'),\\[0.2cm]
\{\bar\Psi^{(u)}(\vec\sigma),\bar\Psi^{(u)}(\vec\sigma')\}_{P.B.}=-4i(\bar
u\bar\pi\bar u)\delta^p(\vec\sigma-\vec\sigma'),\\
\end{array}
\end{equation}
where the r.h.s. of (\ref{59}) are the vector $P^{\dot\alpha\alpha}$ and the
tensor $\pi^{\alpha\beta}$ momenta projections on the isotropic (bi)vectors.

The next first-class constraint from the $\Phi-$sector is the constraint
$\tilde T^{(v)}_R$ (\ref{38}), which closes the Poisson brackets (\ref{37})
for the extra $\kappa-$symmetry generator $\Psi^{(v)}_R$. To prove this
observation let us note that $\tilde T^{(v)}_R$ transforms only the
variables from the $\Phi-$sector $x_{\alpha\dot\alpha}$, $z_{\alpha\beta}$,
$\bar z_{\dot\alpha\dot\beta}$
\begin{equation}\label{60}
\delta_{\tilde T^{(v)}_R}x_{\alpha\dot\alpha}=\epsilon_{\tilde
T^{(v)}_R}v_\alpha\bar v_{\dot\alpha},\quad\delta_{\tilde
T^{(v)}_R}z_{\alpha\beta}=\epsilon_{\tilde T^{(v)}_R}v_\alpha
v_\beta,\quad\delta_{\tilde T^{(v)}_R}\bar
z_{\dot\alpha\dot\beta}=\epsilon_{\tilde T^{(v)}_R}\bar v_{\dot\alpha}\bar
v_{\dot\beta}.
\end{equation}
Using the transformations (\ref{60}) we find
\begin{equation}\label{61}
\delta_{\tilde T^{(v)}_R}S_p=\int d\tau
d^p\sigma\rho^\mu\partial_\mu\epsilon_{\tilde T^{(v)}_R}[(u^\alpha
v_\alpha)^2-1]=0
\end{equation}
and conclude that this transformation is a local symmetry of the
brane action (\ref{14}). The symmetry transformation (\ref{60})
describes local shifts of $x_m$ and $z_{mn}$ coordinates along the
second light-like direction $(v\sigma_m\bar v)$ formed by the dyad
$v_\alpha$. But, unlike the light-like shifts (\ref{48}),
(\ref{50}), the shifts (\ref{60}) are admissible only due to the mutual
cancellation between the $x$ and $z$ contributions into the action
variation (\ref{61}).

This result hints that shifts in the directions transversal to the light-like
ones $u_\alpha\bar u_{\dot\alpha}$, $v_\alpha\bar v_{\dot\alpha}$ and
$u_\alpha u_\beta$, $v_\alpha v_\beta$ may be the symmetries of the action
(\ref{14}) too. To answer this question let us remind that the real basic orts
$m^{(\pm)}_n$ of the local tetrade which are orthogonal to the real
light-like orts $n^{(\pm)}_n$ are given by \cite{zli2}, \cite{GZ}
\begin{equation}\label{62}
m^{(+)}_{\alpha\dot\alpha}=u_\alpha\bar v_{\dot\alpha}+v_\alpha\bar
u_{\dot\alpha},\quad m^{(-)}_{\alpha\dot\alpha}=i(u_\alpha\bar
v_{\dot\alpha}-v_\alpha\bar u_{\dot\alpha}).
\end{equation}
The local shifts of the $x$-coordinates in the transverse directions
$m^{(\pm)}_{\alpha\dot\alpha}$
\begin{equation}\label{63}
\delta_{\Phi^{(+)}}x_{\alpha\dot\alpha}=\epsilon_{\Phi^{(+)}}m^{(+)}_{\alpha\dot\alpha},\quad\delta_{\Phi^{(-)}}x_{\alpha\dot\alpha}=\epsilon_{\Phi^{(-)}}m^{(-)}_{\alpha\dot\alpha}
\end{equation}
generated by the constraints $\Phi^{(\pm)}$
\begin{equation}
\Phi^{(+)}\equiv\Phi^{\dot\alpha\alpha}m^{(+)}_{\alpha\dot\alpha}\approx0,\quad\Phi^{(-)}\equiv\Phi^{\dot\alpha\alpha}m^{(-)}_{\alpha\dot\alpha}\approx0
\end{equation}
change the action (\ref{14})
\begin{equation}\label{65}
\begin{array}{c}
\delta_{\Phi^{(+)}}S_p=\int d\tau
d^p\sigma\rho^\mu\epsilon_{\Phi^{(+)}}(u^\alpha\partial_\mu u_\alpha+\bar
u^{\dot\alpha}\partial_\mu\bar u_{\dot\alpha}),\\[0.2cm]
\delta_{\Phi^{(-)}}S_p=i\int d\tau
d^p\sigma\rho^\mu\epsilon_{\Phi^{(-)}}(u^\alpha\partial_\mu u_\alpha-\bar
u^{\dot\alpha}\partial_\mu\bar u_{\dot\alpha}).
\end{array}
\end{equation}
However, these variations may be compensated by the corresponding
transformations of the TCC coordinates $z_{\alpha\beta}$
\begin{equation}\label{65'}
\begin{array}{c}
\delta_{T^{(+)}}z_{\alpha\beta}=\epsilon_{\Phi^{(+)}}u_{\{\alpha}v_{\beta\}},\quad\delta_{T^{(+)}}\bar
z_{\dot\alpha\dot\beta}=\epsilon_{\Phi^{(+)}}\bar u_{\{\dot\alpha}\bar
v_{\dot\beta\}};\\[0.2cm]
\delta_{T^{(-)}}z_{\alpha\beta}=i\epsilon_{\Phi^{(-)}}u_{\{\alpha}v_{\beta\}},\quad\delta_{T^{(-)}}\bar
z_{\dot\alpha\dot\beta}=-i\epsilon_{\Phi^{(-)}}\bar u_{\{\dot\alpha}\bar
v_{\dot\beta\}}
\end{array}
\end{equation}
generated by the constraints $T^{(+)}$ and $T^{(-)}$
\begin{equation}\label{66}
\begin{array}{c}
 T^{(+)}\equiv\Phi^{\alpha\beta}u_{\{\alpha}v_{\beta\}}+\bar\Phi^{\dot\alpha\dot\beta}\bar u_{\{\dot\alpha}\bar v_{\dot\beta\}}=\Phi^{\alpha\beta}u_{\alpha}v_{\beta}+\bar\Phi^{\dot\alpha\dot\beta}\bar u_{\dot\alpha}\bar v_{\dot\beta}\approx0,\\[0.2cm]
 T^{(-)}\equiv i[\Phi^{\alpha\beta}u_{\{\alpha}v_{\beta\}}-\bar\Phi^{\dot\alpha\dot\beta}\bar u_{\{\dot\alpha}\bar v_{\dot\beta\}}]=i[\Phi^{\alpha\beta}u_{\alpha}v_{\beta}-\bar\Phi^{\dot\alpha\dot\beta}\bar u_{\dot\alpha}\bar v_{\dot\beta}]\approx0,
\end{array}
\end{equation}
where $u_{\{\alpha}v_{\beta\}}=\frac12(u_\alpha v_\beta+u_\beta v_\alpha)$.
In fact, the  variations of the action (\ref{14}) generated by  the doubled  constraints $2T^{(+)}$ and $2T^{(-)}$
\begin{equation}\label{67}
\begin{array}{c}
\delta_{2T^{(+)}}S=-\int d\tau
d^p\sigma\rho^\mu\epsilon_{\Phi^{(+)}}(u^\alpha\partial_\mu u_\alpha+\bar
u^{\dot\alpha}\partial_\mu\bar u_{\dot\alpha}),\\[0.2cm]
\delta_{2T^{(-)}}S=-i\int d\tau
d^p\sigma\rho^\mu\epsilon_{\Phi^{(-)}}(u^\alpha\partial_\mu u_\alpha-\bar
u^{\dot\alpha}\partial_\mu\bar u_{\dot\alpha})
\end{array}
\end{equation}
 exactly compensate the variations (\ref{65}).
It proves that the
two real constraints $\tilde T^{(\pm)}$ belong to the first class
\begin{equation}\label{68}
\begin{array}{c}
\tilde T^{(+)}\equiv \Phi^{(+)}+2T^{(+)} =\Phi^{\dot\alpha\alpha}(u_\alpha\bar
v_{\dot\alpha}+v_\alpha\bar
u_{\dot\alpha})+2\left(\Phi^{\alpha\beta}u_{\{\alpha}v_{\beta\}}+\bar\Phi^{\dot\alpha\dot\beta}u_{\{\dot\alpha}v_{\dot\beta\}}\right)\approx0,\\
[0.2cm]
\tilde T^{(-)}\equiv \Phi^{(-)}+2T^{(-)}=i\left[\Phi^{\dot\alpha\alpha}(u_\alpha\bar
v_{\dot\alpha}-v_\alpha\bar
u_{\dot\alpha})+2\left(\Phi^{\alpha\beta}u_{\{\alpha}v_{\beta\}}-\bar\Phi^{\dot\alpha\dot\beta}u_{\{\dot\alpha}v_{\dot\beta\}}\right)\right]\approx0.
\end{array}
\end{equation}
Thus, we constructed six real bosonic constraints $\Phi^{(u)}$, $T^{(u)}$,
$\bar T^{(u)}$ (\ref{35}), $\tilde T^{(v)}_R$ (\ref{38}) and $\tilde
T^{(\pm)}$ (\ref{68}) belonging to the first class out of the ten
primary constraints (\ref{23}) of the $\Phi-$sector. Further we find
additional first-class constraints that are certain linear combinations of
the constraints from the $\Phi$-, $(u,v)$- and $\rho$-sectors.

\section{Dyad sector: $2\oplus8-$splitting and first-class
constraints}

By analogy with the $\Psi$- and $\Phi$-sectors one can assume that the
first-class constraints from the $(u,v)$-sector describe the local
symmetries related to dyad shifts along themselves. The shifts of $v_\alpha$
  along $u_\alpha$
\begin{equation}\label{69}
\delta_\epsilon v_\alpha=\epsilon u_\alpha,\quad\delta_\epsilon\bar
v_{\dot\alpha}=\bar\epsilon\bar u_{\dot\alpha}
\end{equation}
are generated by the constraints $P^{(u)}_v$ and $\bar P^{(u)}_v$
\begin{equation}\label{70}
P^{(u)}_v\equiv P^\alpha_v u_\alpha\approx0,\quad\bar P^{(u)}_v\equiv\bar
P^{\dot\alpha}_v\bar u_{\dot\alpha}\approx0.
\end{equation}
These shifts supply obvious local symmetry of the action (\ref{14}) since
$S_p$ does not depend on $v_\alpha$. The same is true for the primary
constraints from $\Phi-$, $\Psi-$ and $\rho-$sectors
\begin{equation}
\{P^{(u)}_v(\vec\sigma),\Phi(\vec\sigma')\}_{P.B.}=0,\quad\{P^{(u)}_v(\vec\sigma),\Psi(\vec\sigma)\}_{P.B.}=0,
\quad\{P^{(u)}_v(\vec\sigma),P^{(\rho)}_{\mu}(\vec\sigma')\}_{P.B.}=0.
\end{equation}
Moreover, these shifts do not change the $\Xi$ and $\bar\Xi$ constraints
that depend on $v_\alpha$ and $\bar v_{\dot\alpha}$
\begin{equation}
\{P^{(u)}_v(\vec\sigma),\Xi(\vec\sigma')\}_{P.B.}=0,\quad\{\bar P^{(u)}_v(\vec\sigma),\bar\Xi(\vec\sigma')\}_{P.B.}=0
\end{equation}
as it follows from the canonical relations
\begin{equation}
\{P^\alpha_v(\vec\sigma),v_\beta(\vec\sigma')\}_{P.B.}=\delta^\alpha_\beta\delta^p(\vec\sigma-\vec\sigma'),\quad\{\bar
P^{\dot\alpha}_v(\vec\sigma),\bar
v_{\dot\beta}(\vec\sigma')\}_{P.B.}=\delta^{\dot\alpha}_{\dot\beta}\delta^p(\vec\sigma-\vec\sigma')
\end{equation}
and their complex conjugate.
Therefore, the two real constraints (\ref{70}) are the first-class
constraints, and they have zero Poisson brackets between themselves
\begin{equation}
\{P^{(u)}_v(\vec\sigma),P^{(u)}_v(\vec\sigma')\}_{P.B.}=0,\quad\{P^{(u)}_v(\vec\sigma),\bar
P^{(u)}_v(\vec\sigma')\}_{P.B.}=0.
\end{equation}

However, the $v-$shifts along themselves generated by the constraints
$P^{(v)}_v$ and $\bar P^{(v)}_v$
\begin{equation}\label{75}
P^{(v)}_v\equiv P^\alpha_v v_\alpha\approx0,\quad\bar P^{(v)}_v\equiv\bar
P^{\dot\alpha}_v\bar v_{\dot\alpha}\approx0
\end{equation}
do not preserve the constraints $\Xi$ and $\bar\Xi$
\begin{equation}
\{P^{(v)}_v(\vec\sigma),\Xi(\vec\sigma')\}_{P.B.}=(1+\Xi)\delta^p(\vec\sigma-\vec\sigma'),\quad\{\bar
P^{(v)}_v(\vec\sigma),\bar\Xi(\vec\sigma')\}_{P.B.}=(1+\bar\Xi)\delta^p(\vec\sigma-\vec\sigma').
\end{equation}
As a result, the constraints (\ref{75}), as well as the constraints
$P^{(u)}_u$ and $\bar P^{(u)}_u$
\begin{equation}\label{77}
P^{(u)}_u\equiv P^\alpha_u u_\alpha\approx0,\quad\bar P^{(u)}_u\equiv\bar
P^{\dot\alpha}_u\bar u_{\dot\alpha}\approx0
\end{equation}
are not the first-class ones.

Moreover, the $u_\alpha-$shifts along $v_\alpha$, generated by the
constraints $P^{(v)}_u$ and $\bar P^{(v)}_u$
\begin{equation}\label{79}
P^{(v)}_u\equiv P^\alpha_u v_\alpha\approx0,\quad\bar P^{(v)}_u\equiv\bar
P^{\dot\alpha}_u\bar v_{\dot\alpha}\approx0
\end{equation}
which also have zero Poisson brackets with $\Xi$ and $\bar\Xi$
\begin{equation}
\{P^{(v)}_u(\vec\sigma),\Xi(\vec\sigma')\}_{P.B.}=0,\quad\{\bar P^{(v)}_u(\vec\sigma),\bar\Xi(\vec\sigma')\}_{P.B.}=0,
\end{equation}
are not the symmetries of $S_p$.
 These constraints have also nonzero Poisson
brackets with the $\Phi-$sector constraints
\begin{equation}\label{81}
\begin{array}{c}
\{\Phi^{\dot\alpha\alpha}(\vec\sigma),P^\beta_{u}(\vec\sigma')\}_{P.B.}=\varepsilon^{\alpha\beta}\rho^\tau\bar
u^{\dot\alpha}\delta^p(\vec\sigma-\vec\sigma'),\\[0.2cm]
\{\Phi^{\alpha\beta}(\vec\sigma),P^\gamma_{u}(\vec\sigma')\}_{P.B.}=-\frac12\rho^\tau(\varepsilon^{\alpha\gamma}u^{\beta}+\varepsilon^{\beta\gamma}u^{\alpha})\delta^p(\vec\sigma-\vec\sigma'),
\end{array}
\end{equation}
where the canonical P.B. relations
\begin{equation}
\{u^\alpha(\vec\sigma),P^\beta_u(\vec\sigma')\}_{P.B.}=-\varepsilon^{\alpha\beta}\delta^p(\vec\sigma-\vec\sigma')
\end{equation}
were used.
After multiplication of (\ref{81}) by $v_\beta(\tau,\vec\sigma')$ and $v_\gamma(\tau,\vec\sigma')$, respectively,
we find
\begin{equation}\label{83}
\begin{array}{c}
\{\Phi^{\dot\alpha\alpha}(\vec\sigma),P^{(v)}_{u}(\vec\sigma')\}_{P.B.}=\rho^\tau
v^\alpha\bar u^{\dot\alpha}\delta^p(\vec\sigma-\vec\sigma'),\\[0.2cm]
\{\Phi^{\alpha\beta}(\vec\sigma),P^{(v)}_{u}(\vec\sigma')\}_{P.B.}=-\frac12\rho^\tau(v^{\alpha}u^{\beta}+v^{\beta}u^{\alpha})\delta^p(\vec\sigma-\vec\sigma'),
\end{array}
\end{equation}
where the r.h.s. of (\ref{83}) do not belong to the primary constraints. So,
we resume that the $(u,v)-$sector itself includes only two real
first-class constraints,    but, as we shall see below, some of the considered
shifts may be compensated by the transformations from the $\rho-$sector.

\section{$\rho-$sector: $p\oplus1-$splitting and first-class
constraints}

The $\rho-$sector of the $p-$brane constraints (\ref{26}) includes $p$
constraints of the first-class
\begin{equation}\label{84}
P^{(\rho)}_{M}\approx0,\quad M=(1,...,p).
\end{equation}
It follows from the observation that the corresponding canonically
conjugate variables $\rho^{M}$
\begin{equation}\label{85}
\{P^{(\rho)}_{M}(\vec\sigma),\rho^{N}(\vec\sigma')\}_{P.B.}=\delta^{N}_{M}\delta^p(\vec\sigma-\vec\sigma')
\end{equation}
do not enter the primary constraints (\ref{23})-(\ref{26}) and, consequently,
$P^{(\rho)}_{M}$ have zero Poisson brackets with all these
constraints. $P^{(\rho)}_{M}$ constraints correspond to redefinition of $p$ space components $\rho^{M}$ of the
$(p+1)-$dimensional world-volume vector density $\rho^\mu(\tau,\vec\sigma)$
\begin{equation}\label{86}
\delta_\epsilon\rho^{M}=\epsilon^{M}(\tau,\vec\sigma).
\end{equation}
The transformations (\ref{86}) are new local symmetries of the action $S_p$
due to arbitrariness in the definition of $\rho^{M}$. On the
other hand, the world-volume time component $\rho^\tau$ enters in the $\Phi-$sector
constraints and the remaining component $P^{(\rho)}_\tau$ from the
$\rho-$sector has nonzero Poisson brackets with this sector constraints
\begin{equation}
\begin{array}{c}
\{P^{(\rho)}_\tau(\vec\sigma),\Phi^{\dot\alpha\alpha}(\vec\sigma')\}_{P.B.}=-u^\alpha\bar
u^{\dot\alpha}\delta^p(\vec\sigma-\vec\sigma'),\\[0.2cm]
\{P^{(\rho)}_\tau(\vec\sigma),\Phi^{\alpha\beta}(\vec\sigma')\}_{P.B.}=\frac12
u^\alpha u^{\beta}\delta^p(\vec\sigma-\vec\sigma'),\\[0.2cm]
\{P^{(\rho)}_\tau(\vec\sigma),\bar\Phi^{\dot\alpha\dot\beta}(\vec\sigma')\}_{P.B.}=\frac12\bar
u^{\dot\alpha}\bar u^{\dot\beta}\delta^p(\vec\sigma-\vec\sigma').
\end{array}
\end{equation}
However, the transformation of $\rho^\tau$
\begin{equation}\label{88}
\delta_\epsilon\rho^{\tau}=\epsilon^{\tau}(\tau,\vec\sigma)
\end{equation}
 generated by the constraint $P^{(\rho)}_\tau$, which is not the first-class one, may be compensated by the corresponding transformation of dyads as we shall show below.

\section{The Weyl symmetry constraint}

Here we find the first-class constraint showing the presence of the local
Weyl symmetry of the brane action. Because the $\Phi$-sector constraints contain the dyad $u_\alpha$, the transformation (\ref{88}) can be compensated by the following transformation of $u_\alpha$
\begin{equation}
\delta_\epsilon
u_\alpha=-\frac{\epsilon^\tau}{2\rho^\tau}u_\alpha,\quad\delta_\epsilon\bar
u_{\dot\alpha}=-\frac{\epsilon^\tau}{2\rho^\tau}\bar u_{\dot\alpha},
\end{equation}
 generated by the constraint $\Delta^{(u)}$
\begin{equation}\label{89}
\Delta^{(u)}\equiv P^\alpha_u u_\alpha+\bar P^{\dot\alpha}_u\bar
u_{\dot\alpha}-2\rho^\tau P^{(\rho)}_\tau\approx0.
\end{equation}
However, the new constraint (\ref{89}) has nonzero Poisson brackets with $\Xi$
and $\bar\Xi$
\begin{equation}\label{90}
\{\Delta^{(u)}(\vec\sigma),\Xi(\vec\sigma')\}_{P.B.}=(1+\Xi)\delta^p(\vec\sigma-\vec\sigma'),\quad\{\Delta^{(u)}(\vec\sigma),\bar\Xi(\vec\sigma')\}_{P.B.}=(1+\bar\Xi)\delta^p(\vec\sigma-\vec\sigma').
\end{equation}
This noncommutativity can be easily corrected by the compensating
transformation of $v_\alpha$
\begin{equation}\label{91}
\delta_\epsilon
v_\alpha=\frac{\epsilon^\tau}{2\rho^\tau}v_\alpha,\quad\delta_\epsilon\bar
v_{\dot\alpha}=\frac{\epsilon^\tau}{2\rho^\tau}\bar v_{\dot\alpha}.
 \end{equation}
It implies the generalization of $\Delta^{(u)}$ (\ref{89}) into the new constraint $\Delta'_W$
\begin{equation}\label{92}
\Delta'_W\equiv(P^\alpha_u u_\alpha+\bar P^{\dot\alpha}_u\bar
u_{\dot\alpha})-(P^\alpha_v v_\alpha+\bar P^{\dot\alpha}_v\bar
v_{\dot\alpha})-2\rho^\tau P^{(\rho)}_\tau\approx0
\end{equation}
which  has zero Poisson brackets with the $\Phi-$sector. Moreover, the Poisson
brackets of $\Delta'_W$ with the primary constraints (\ref{23})-(\ref{26})
are equal to zero on the constraint surface. These properties of the
constraint $\Delta'_W$ will be preserved after addition of the first-class
constraints (\ref{84}) which transform $\Delta'_W$ to
\begin{equation}\label{93}
\Delta_W\equiv(P^\alpha_u u_\alpha+\bar P^{\dot\alpha}_u\bar
u_{\dot\alpha})-(P^\alpha_v v_\alpha+\bar P^{\dot\alpha}_v\bar
v_{\dot\alpha})-2\rho^\mu P^{(\rho)}_\mu\approx0.
\end{equation}
The local transformation generated by $\Delta_W$ is the dilation affecting only the auxiliary fields $u_\alpha$, $v_\alpha$ and $\rho^\mu$
\begin{equation}\label{94}
\begin{array}{c}
\rho'^{\mu}=e^{-2\Lambda}\rho^\mu,\; u'_\alpha=e^\Lambda u_\alpha,\;
v'_\alpha=e^{-\Lambda} v_\alpha,\\
x'_{\alpha\dot\alpha}=x_{\alpha\dot\alpha},\
z'_{\alpha\beta}=z_{\alpha\beta},\ \theta'_\alpha= \theta_\alpha.
\end{array}
\end{equation}
The transformation (\ref{94}) is identified with the Weyl symmetry of the $p-$brane action. From the string point of view, the Weyl invariants $\rho^\mu u_\alpha\bar u_{\dot\alpha}$ and $\rho^\mu u_\alpha u_\beta$ constructed from auxiliary world-volume fields are similar to the conventional Weyl invariant of tensile
string
\begin{equation}\label{95}
\sqrt{-g}g^{\mu\nu}\Leftrightarrow\rho^\mu u_\alpha\bar u_{\dot\alpha},
\end{equation}
but here (\ref{94}) is the symmetry the tensionless super $p$-brane action.

The transformations (\ref{86}) can be used for the gauge fixing
\begin{equation}\label{96}
\rho^{M}=0.
\end{equation}
Then the Weyl symmetry may be used to fix $\rho^\tau$ by the gauge condition
\begin{equation}\label{97}
\dot\rho^\tau(\tau,\vec\sigma)=0,
\end{equation}
or $\rho^\tau=\rho^0=\mbox{const}$.

Below we shall show that the $\rho^M$ translations (\ref{86}) are supplemented by  reparametrization transformation of the world-volume coordinates $\sigma_{M}$ generated by $p$ first-class constraints formed by intermixing of the primary constraints.

\section{Secondary constraints: $p$ space-like reparametrizations}

To find the first-class constraints associated with the reparametrizations
of $p$ world-volume coordinates $\sigma_{M}$, let us consider the
canonical Hamiltonian for $p-$brane action (\ref{14}). \\
Using the standard
definition of the canonical Hamiltonian density
\begin{equation}\label{104}
H_0=\dot\EuScript Q_{\mathfrak M}\EuScript P^{\mathfrak M}-L=\dot
x_{\alpha\dot\alpha}P^{\dot\alpha\alpha}+\dot
z_{\alpha\beta}\pi^{\alpha\beta}+\dot{\bar
z}_{\dot\alpha\dot\beta}\bar\pi^{\dot\alpha\dot\beta}+\dot\theta_\alpha\pi^\alpha+\dot{\bar\theta}_{\dot\alpha}\bar\pi^{\dot\alpha}+(\dot
u_\alpha P^\alpha_u+\dot v_\alpha P^\alpha_v+c.c.)+\dot\rho^\mu
P^{(\rho)}_\mu-L
\end{equation}
and the $p-$brane Lagrangian (\ref{14}) we find
\begin{equation}\label{105}
H_0\approx-\rho^{M}[u^\alpha\omega_{M\alpha\dot\alpha}\bar
u^{\dot\alpha}+\frac12
u^\alpha\omega_{M\alpha\beta}u^\beta+\frac12\bar
u^{\dot\alpha}\bar\omega_{M\dot\alpha\dot\beta}\bar u^{\dot\beta}],
\end{equation}
where we omitted the momenta from the $(u,v)$- and $\rho$-sectors equal to zero. Taking into account the definitions (\ref{23}) one obtains
\begin{equation}\label{106}
\begin{array}{rl}
H_0\approx&-\frac{\rho^{M}}{\rho^\tau}\left[-(\Phi^{\dot\alpha\alpha}-P^{\dot\alpha\alpha})\omega_{M\alpha\dot\alpha}+(\Phi^{\alpha\beta}-\pi^{\alpha\beta})\omega_{M\alpha\beta}+(\bar\Phi^{\dot\alpha\dot\beta}-\bar\pi^{\dot\alpha\dot\beta})\bar\omega_{M\dot\alpha\dot\beta}\right]\\[0.2cm]
\approx&-\frac{\rho^{M}}{\rho^\tau}\left[P^{\dot\alpha\alpha}\omega_{M\alpha\dot\alpha}-\pi^{\alpha\beta}\omega_{M\alpha\beta}-\bar\pi^{\dot\alpha\dot\beta}\bar\omega_{M\dot\alpha\dot\beta}\right].
\end{array}
\end{equation}
Due to the $\Psi-$constraints (\ref{24}) there is the equality
\begin{equation}\label{107}
\begin{array}{rl}
l^{(\Phi,\Psi)}_{M}\equiv&P^{\dot\alpha\alpha}\partial_{M}x_{\alpha\dot\alpha}+\pi^{\alpha\beta}\partial_{M}z_{\alpha\beta}+\bar\pi^{\dot\alpha\dot\beta}\partial_{M}\bar
z_{\dot\alpha\dot\beta}+\partial_{M}\theta_\alpha\pi^\alpha+\partial_{M}\bar\theta_{\dot\alpha}\bar\pi^{\dot\alpha}
\\[0.2cm]
=&P^{\dot\alpha\alpha}\omega_{M\alpha\dot\alpha}-\pi^{\alpha\beta}\omega_{M\alpha\beta}-\bar\pi^{\dot\alpha\dot\beta}\bar\omega_{M\dot\alpha\dot\beta}+\partial_{M}\theta_\alpha\Psi^\alpha+\partial_{M}\bar\theta_{\dot\alpha}\bar\Psi^{\dot\alpha},
\end{array}
\end{equation}
where $l^{(\Phi,\Psi)}_{M}$ is the reparametrization generator of the
$\Phi$- and $\Psi$-sector coordinates.\\
 The total reparametrization generator
$L_{M}$ is connected with $l^{(\Phi,\Psi)}_{M}$ by the equality
\begin{equation}\label{108}
L_{M}=l^{(\Phi,\Psi)}_{M}+l^{(u,v,\rho)}_{M},
\end{equation}
where $l^{(u,v,\rho)}_{M}$ is the reparametrization generator for the coordinates $u$,
$v$, $\rho$
\begin{equation}\label{109}
l^{(u,v,\rho)}_{M}\equiv(P^\alpha_u\partial_M u_\alpha+P^\alpha_v\partial_M
v_\alpha)+(\bar P^{\dot\alpha}_u\partial_M{\bar u}_{\dot\alpha}+\bar
P^{\dot\alpha}_v\partial_M{\bar v}_{\dot\alpha})-\partial_M
P^{(\rho)}_{\nu}\rho^{\nu}\approx0.
\end{equation}
It follows from Eqs.(\ref{107})-(\ref{109}) that $H_0$ (\ref{105}) may be
presented in the form
\begin{equation}\label{110}
H_0\approx-\frac{\rho^{M}}{\rho^\tau}L_{M},\quad(M=1,...,p).
\end{equation}
As a result of the  Dirac selfconsistency condition for the constraints (\ref{84})
\begin{equation}\label{111}
\dot P^{(\rho)}_{M}=\int
d^p\sigma'\{H_T(\vec\sigma'),P^{(\rho)}_{M}(\vec\sigma)\}_{P.B.}=\int
d^p\sigma'\{H_0(\vec\sigma'),P^{(\rho)}_{M}(\vec\sigma)\}_{P.B.}=\frac{1}{\rho^\tau}L_{M}\approx0
\end{equation}
we find $p$ new constraints $L_{M}$
\begin{equation}\label{112}
\begin{array}{rl}
L_{M}&=P^{\dot\alpha\alpha}\omega_{M\alpha\dot\alpha}-\pi^{\alpha\beta}\omega_{M\alpha\beta}-\bar\pi^{\dot\alpha\dot\beta}\bar\omega_{M\dot\alpha\dot\beta}+\partial_{M}\theta_\alpha\Psi^\alpha+\partial_{M}\bar\theta_{\dot\alpha}\bar\Psi^{\dot\alpha}\\[0.2cm]
&+(P^\alpha_u\partial_M u_\alpha+P^\alpha_v\partial_M v_\alpha)+(\bar
P^{\dot\alpha}_u\partial_M{\bar u}_{\dot\alpha}+\bar P^{\dot\alpha}_v\partial_M{\bar
v}_{\dot\alpha})-\partial_M P^{(\rho)}_{\nu}\rho^{\nu}\approx0.
\end{array}
\end{equation}
One can check that $L_{M}$ have Poisson brackets with the primary constraints from all sectors weakly equal to zero. Thus, these
constraints realize the $\sigma_{M}$ transformations of the
reparametrization invariance of the brane action. These secondary
constraints complete realization of the Dirac procedure of the first-class
constraints construction. The remaining time-like $\tau$-reparametrization constraint is not independent and is constructed from the constraints $\Phi^{(u)}$, $T^{(u)}$ (\ref{35}) for $ x_{\alpha\dot\alpha}$, $z_{\alpha\beta}$ and some algebraic combinations of other first-class constraints for the remaining generalized coordinates.

\section{$Y^\Lambda$ supertwistor as an invariant of local symmetries}

Here we shall show that $(8_B+1_F)$ real components of the supertwistor
$Y^\Lambda=(iU^a,\tilde Y_a,\tilde\eta)$ in $4-$dimensional Minkowski space
extended by 6 TCC coordinates are the invariants of $(8_B+3_F)$ local symmetries generated by the first-class constraints from the $\Phi$-, $\Psi$- and dyad
sectors. We prove that the transition from the original representation of the action (\ref{1}) (or (\ref{14})) in terms of $(16_B+4_F)$ variables
$(x,z,u,v;\theta)$ to the $Y^\Lambda$ supertwistor representation (\ref{6}), including $(8_B+1_F)$ components, preserves all the local
symmetries of $S_p$ (\ref{1}). It will follow from the observation that $Y^\Lambda$ forms a representation of the Weyl symmetry (\ref{94}), the space-like
reparametrizations (\ref{112}) and its invariance under the $\rho^{M}-$shifts (\ref{86}). All these $(2p+1)$ symmetries remain local symmetries of the supertwistor representation (\ref{6}).

Starting the proof we observe that the supertwistor component
$U_{a}= {u_\alpha \choose \bar u^{\dot\alpha}}$ is  a trivial invariant of all the 11 symmetries generated by the $\Psi$-, $\Phi$-
and $(u,v)$-sectors. The fermionic component
$\tilde\eta=-2i(U^a\theta_a)=-2i(u^\alpha\theta_\alpha+\bar
u_{\dot\alpha}\bar\theta^{\dot\alpha})$ is not transformed under symmetry
transformations from the $\Phi-$ and $(u,v)-$sectors. Moreover, its
invariance under the $\kappa-$symmetry generated by $\Psi^{(u)}$,
$\bar\Psi^{(u)}$ and $\Psi^{(v)}_R$ (\ref{43}) follows from the relations
\begin{equation}\label{113}
\delta_{\kappa}\tilde\eta={\textstyle\frac{2}{i}}\kappa u^\alpha
u_\alpha=0,\quad\delta_{\bar\kappa}\tilde\eta=-{\textstyle\frac{2}{i}}\bar\kappa\bar
u^{\dot\alpha}\bar
u_{\dot\alpha}=0,\quad\delta_{\tilde\kappa_R}\tilde\eta={\textstyle\frac{2}{i}}\tilde\kappa_R
[u^\alpha v_\alpha+\bar u_{\dot\alpha}\bar v^{\dot\alpha}]\approx 0.
\end{equation}

Thus, it remains to prove the invariance of the $\tilde Y_a$ components of the
supertwistor
\begin{equation}\label{114}
i\tilde Y_a=Y_{ab}U^b-\tilde\eta\theta_a={-z_{\alpha\beta}u^\beta+x_{\alpha\dot\alpha}\bar
u^{\dot\alpha}-\tilde\eta\theta_\alpha\choose\tilde
x^{\dot\alpha\alpha}u_\alpha-\bar z^{\dot\alpha\dot\beta}\bar
u_{\dot\beta}-\tilde\eta\bar\theta^{\dot\alpha}}.
\end{equation}
Due to the invariance of $U_a$ and $\tilde\eta$ variables the variation of
$\tilde Y_a$ is given by
\begin{equation}\label{115}
i\delta\tilde Y_a=\delta Y_{ab}U^b-\tilde\eta\delta\theta_a= {-\delta
z_{\alpha\beta}u^\beta+\delta x_{\alpha\dot\alpha}\bar
u^{\dot\alpha}-\tilde\eta\delta\theta_\alpha\choose\delta\tilde
x^{\dot\alpha\alpha}u_\alpha-\delta\bar z^{\dot\alpha\dot\beta}\bar
u_{\dot\beta}-\tilde\eta\delta\bar\theta^{\dot\alpha}}
\end{equation}
and it is enough to verify the invariance of the upper Weyl component
$iy_\alpha$. Using the relations (\ref{43}) we find that
\begin{equation}\label{116}
i\delta_{\kappa}y_\alpha=-2i\kappa u_\alpha[u^\beta\theta_\beta-\bar
u^{\dot\beta}\bar\theta_{\dot\beta}-{\textstyle\frac{i}{2}}\tilde\eta]=0,\
i\delta_{\tilde\kappa_R}y_\alpha=-2i\tilde\kappa_R
v_\alpha[u^\beta\theta_\beta-\bar
u^{\dot\beta}\bar\theta_{\dot\beta}-{\textstyle\frac{i}{2}}\tilde\eta]=0.
\end{equation}
So, we confirmed the invariant character of $\tilde Y_a$ under the
$\kappa-$symmetry transformations.

 The invariance of $\tilde Y_a$ under the
symmetry transformations (\ref{69}) from the $(u,v)-$sector is evident,
because $\tilde Y_a$ does not include $v_\alpha$.

The transformations of $\tilde Y_a$ under the symmetries of the $\Phi-$sector are defined by
\begin{equation}\label{117}
i\delta\tilde Y_a={-\delta z_{\alpha\beta}u^\beta+\delta
x_{\alpha\dot\alpha}\bar u^{\dot\alpha}\choose\delta\tilde
x^{\dot\alpha\alpha}u_\alpha-\delta\bar z^{\dot\alpha\dot\beta}\bar
u_{\dot\beta}},
\end{equation}
because $\theta_a$ and $\tilde\eta$ are invariants of these symmetries.
 The invariance of $\tilde Y_a$ under the symmetries (\ref{47}) and (\ref{50}) is
evident because of the relations
\begin{equation}\label{118}
\delta_{\Phi^{(u)}} x_{\alpha\dot\alpha}\bar
u^{\dot\alpha}=\epsilon_{\Phi^{(u)}}u_\alpha(\bar u_{\dot\alpha}\bar
u^{\dot\alpha})=0,\quad
\delta_{T^{(u)}}z_{\alpha\beta}u^{\beta}=\epsilon_{T^{(u)}}u_\alpha(u_{\beta}u^{\beta})=0
\end{equation}
and their complex conjugate.

The invariance under the symmetry (\ref{60}) follows from the cancellation between the
$x$ and $z$ contributions
\begin{equation}\label{119}
\delta_{\tilde T^{(v)}_R} x_{\alpha\dot\alpha}\bar u^{\dot\alpha}-
\delta_{\tilde T^{(v)}_R}z_{\alpha\beta}u^\beta
\approx
\epsilon_{\tilde T^{(v)}_R} v_\alpha-\epsilon_{\tilde T^{(v)}_R} v_\alpha=0.
\end{equation}
Analogous cancellations also take place with respect to the
transformations (\ref{63}), (\ref{65'})
\begin{equation}\begin{array}{c}\label{120}
\delta_{\tilde T^{(+)}} x_{\alpha\dot\alpha}\bar u^{\dot\alpha}-\delta_{\tilde T^{(+)}}
z_{\alpha\beta}u^\beta=\epsilon_{\tilde T^{(+)}} [m^{(+)}_{\alpha\dot\alpha}\bar
u^{\dot\alpha}- u_{\{\alpha}v_{\beta\}}u^\beta]\approx 0,\\
\delta_{\tilde T^{(-)}} x_{\alpha\dot\alpha}\bar u^{\dot\alpha}-\delta_{\tilde T^{(-)}}
z_{\alpha\beta}u^\beta=\epsilon_{\tilde T^{(-)}} [m^{(-)}_{\alpha\dot\alpha}\bar
u^{\dot\alpha}-iu_{\{\alpha}v_{\beta\}}u^\beta]\approx 0
\end{array}
\end{equation}
and their complex conjugate. It completes the proof of invariant character
of $Y^\Lambda$ under $(8_B+3_F)$ local symmetries generated by the
$\Psi$-, $\Phi$- and $(u,v)$-sectors. These symmetries show the pure gauge
character of $(8_B+3_F)$ variables among $(16_B+4_F)$ primary variables
$(x,z,u,v;\theta)$. The transition to the supertwistor representation
including $(8_B+1_F)$ invariant variables $Y^\Lambda$ just encodes these
$(8_B+3_F)$ degrees of freedom
\begin{equation}
(16_B+4_F)-(8_B+3_F)=(8_B+1_F),
\end{equation}
so we get the next scheme of the reduction of the original brane variables:
\begin{equation}\label{121}
\begin{array}{c}
{\kappa-\mbox{symmetry:}}\quad\theta_\alpha,\
\bar\theta_{\dot\alpha}\Rightarrow\tilde\eta,\\
{\mbox{shifts:}}\quad x_{\alpha\dot\alpha}, z_{\alpha\beta}, \bar
z_{\dot\alpha\dot\beta}; v_\alpha,\bar v_{\dot\alpha}\Rightarrow(\tilde
Y_\alpha,\bar{\tilde Y}_{\dot\alpha}).
\end{array}
\end{equation}

Taking into account the Weyl symmetry transformations (\ref{94}) one can
find the Weyl transformation of the supertwistor
\begin{equation}\label{122}
\acute Y^\Sigma=e^{\Lambda(\tau,\vec\sigma)}Y^\Sigma.
\end{equation}
Using this result one verifies that the supertwistor representation
(\ref{6}) of $S_p$ is invariant under the Weyl symmetry. Indeed, the
transformed action (\ref{6}) is
\begin{equation}\label{123}
\begin{array}{rl}
S'_p&=\frac12\int d\tau d^p\sigma\rho^\mu\left[\partial_\mu Y^\Sigma
G_{\Sigma\Xi}Y^{\Xi}-\partial_\mu\Lambda(Y^{\Sigma}G_{\Sigma\Xi}Y^{\Xi})\right]\\[0.2cm]
&=S_p-\frac12\int d\tau
d^p\sigma\rho^\mu\partial_\mu\Lambda(Y^{\Sigma}G_{\Sigma\Xi}Y^{\Xi}),
\end{array}
\end{equation}
but the last term equals zero, because $Y^{\Sigma}G_{\Sigma\Xi}Y^{\Xi}=0$ and
 $S'_p=S_p$. Also, the supertwistor action is invariant under $p$ remaining local  symmetries (\ref{86}) from $\rho-$sector and $p$ world-volume reparametrizations generated by the secondary first-class constraints (\ref{112}).

Thus, we conclude that the strong reduction of the number of variables
during the transition to the symplectic supertwistor representation (\ref{6}) is a consequence of the described local symmetries. The pure gauge degrees of
freedom are eliminated by the change of variables without any gauge fixing.

\section{The Hamiltonian}

Having completed the Dirac prescription of the first-class constraints construction we present the total Hamiltonian density of the super $p$-brane in the form of their linear combination
\begin{equation}\begin{array}{rl}\label{126}
H_T=&\kappa_u\Psi^{(u)}+\bar\kappa_u\bar\Psi^{(u)}+\kappa_R\Psi^{(v)}_R\\
+&a_u\Phi^{(u)}+b_u T^{(u)}+\bar b_u\bar T^{(u)}\\
+&c^{(+)}\tilde T^{(+)}+c^{(-)}\tilde T^{(-)}+c^{(v)}_R\tilde T^{(v)}_R\\
+&e P^{(u)}_v+\bar e \bar P^{(u)}_v+\omega\Delta_W\\
+&f^{M}P^{(\rho)}_{M}+\tilde\rho^{M}L_{M}\approx0,
\end{array}
\end{equation}
where the functions $\kappa$, $a$, $b$, $c$, $e$, $f$, $\omega$ and  $\tilde\rho$  form the set of $(9+2p)_B+3_F$ real Lagrange multipliers.

In the Hamiltonian formalism the second-class constraints are taken into account by the transition to the  Dirac brackets (D.B.) which have to be changed by the (anti)commutators in the quantum dynamics. To construct  the D.B. we need to find  the second-class constraints remaining after the first-class 
constraint separation. This problem will be solved below.

\section{Second-class constraints and the Dirac brackets}

To find  the second-class constraints we shall follow to the applied above projection method \cite{BZ90}, \cite{zli2}, \cite{GZ}. In the $\Psi$-sector only one constraint $\Psi^{(v)}_I$ (\ref{38'}) has remained
 which is the second-class  constraint, because its  P.B.
\begin{equation}\label{126'}
\{\Psi^{(v)}_I(\vec\sigma), \Psi^{(v)}_I(\vec\sigma')\}_{P.B.}=8i[(v_\alpha\Phi^{\alpha\beta}v_{\beta}+
\bar v_{\dot\alpha}\bar\Phi^{\dot\alpha\dot\beta}\bar
v_{\dot\beta}- \bar v_{\dot\beta}\Phi^{\dot\beta\alpha}v_{\alpha})
-2\rho^{\tau}]\delta^p(\vec\sigma-\vec\sigma')
 \end{equation}
are nonzero in the weak sense
\begin{equation}\label{127}
\{\Psi^{(v)}_I(\vec\sigma), \Psi^{(v)}_I(\vec\sigma')\}_{P.B.}\approx
-16i\rho^{\tau}\delta^p(\vec\sigma-\vec\sigma').
\end{equation}

In the  $\Phi$-sector there are four second-class constraints that  may be constructed  either from  the $\Phi^{\alpha\beta}$-subsector alone or from the  $\Phi^{\alpha\beta}$- and  $\Phi^{\dot\alpha\alpha}$-subsectors.
We choose the projections of the three remaining constraints from the  $\Phi^{\dot\alpha\alpha}$-subsector
\begin{equation}\label{128}
\begin{array}{c}
\Phi\equiv\Phi^{\dot\alpha\alpha}(m^{(+)}-im^{(-)})_{\alpha\dot\alpha}=
2\bar v_{\dot\alpha}\Phi^{\dot\alpha\alpha}u_{\alpha} \approx 0 \\[0.2cm]
\bar\Phi\equiv (\Phi)^{*}=\Phi^{\dot\alpha\alpha}(m^{(+)}+im^{(-)})_{\alpha\dot\alpha}=
2\bar u_{\dot\alpha}\Phi^{\dot\alpha\alpha}v_{\alpha}\approx 0 ,
 \\[0.2cm]
\Phi^{(v)}\equiv \bar v_{\dot\alpha}\Phi^{\dot\alpha\alpha}v_{\alpha}\approx 0
\end{array}
\end{equation}
to be the second-class constraints.
Then the  additional fourth  second-class constraint $T^{(v)}_I$ is supplied  by the $\Phi^{\alpha\beta}$-subsector
\begin{equation}\label{129}
T^{(v)}_I\equiv i(T^{(v)} -\bar T^{(v)})=i (v_{\alpha}\Phi^{\alpha\beta}
v_{\beta}- \bar v_{\dot\alpha}\bar\Phi^{\dot\alpha\dot\beta}\bar v_{\dot\beta})
\approx 0.
\end{equation}

The  second-class constraints  (\ref{128}), (\ref{129})  have non-zero P.B. with the constraints  belonging to the dyad and  $\rho$-sectors. Therefore, we shall seek for such linear combinations from these sectors that form canonically conjugate  pairs with the constraints  (\ref{128}), (\ref{129}).
 
Then  we find the following four second-class constraints
\begin{equation}\label{130}
P^{(\rho)}_{\tau}\approx 0 , \quad
\{P^{(\rho)}_{\tau}(\vec\sigma),\Phi^{(v)}(\vec\sigma')\}_{P.B.}
\approx\delta^p(\vec\sigma-\vec\sigma'),
\end{equation}
\begin{equation}\label{131}
\begin{array}{c}
P^{(v)}_{u}\equiv P^{\alpha}_{u}v_{\alpha}\approx 0, \quad
\{ P^{(v)}_{u}(\vec\sigma),\Phi(\vec\sigma') \}_{P.B.}\approx 2\rho^{\tau}\delta^p(\vec\sigma-\vec\sigma'), \\[0.2cm]
\bar P^{(v)}_{u}\equiv \bar P^{\dot\alpha}_{u}\bar v_{\dot\alpha}\approx 0, 
\quad
\{ \bar P^{(v)}_{u}(\vec\sigma),\bar\Phi(\vec\sigma') \}_{P.B.}\approx 2
\rho^{\tau}
\delta^p(\vec\sigma-\vec\sigma')
\end{array}
\end{equation}
and
\begin{equation}\label{132}
\begin{array}{r}
\Delta_{I}\equiv \Delta^{(v)}_{I}- \Delta^{(u)}_{I} \equiv
i(P^{(v)}_{v}- \bar P^{(v)}_{v})-i(P^{(u)}_{u}- \bar P^{(u)}_{u}) \\[0.2cm]
\equiv
i(P^\alpha_v v_\alpha - \bar P^{\dot\alpha}_v\bar v_{\dot\alpha})-
i(P^\alpha_u u_\alpha - \bar P^{\dot\alpha}_u\bar u_{\dot\alpha})\approx0,\\[0.2cm]
\{\Delta_{I}(\vec\sigma), T^{(v)}_{I}(\vec\sigma')\}_{P.B.}
\approx 2\rho^{\tau}\delta^p(\vec\sigma-\vec\sigma').
\end{array}
\end{equation}
The remaining four second-class constraints forming canonically conjugate pairs are supplied  by the constraints belonging to the dyad and $\rho$-sectors
\begin{equation}\label{133}
\begin{array}{c}
\Delta\equiv P^{(u)}_{u}+ P^{(v)}_{v}- \rho^{(\tau)} P^{(\rho)}_\tau -
\frac{i}{2}\Delta_{I}\approx 0, \quad  \Xi\approx 0 , \\[0.2cm]
\{ \Delta(\vec\sigma),\Xi(\vec\sigma')\}_{P.B.}= 2\delta^{p}(\vec\sigma-\vec\sigma')
\end{array}
\end{equation}
and their  complex conjugate. As a result, the P.B's. of 12 bosonic constraints (\ref{128})-(\ref{133}) and one fermionic constraint (\ref{38'}) form the Dirac matrix $\hat{\mathbf C}$ having the  symplectic form
\begin{equation}\label{135}
\hat{\mathbf C}=\begin{array} {l|ccccccccccccc|}
\multicolumn{14}{c}{\hspace*{0.7cm} P^{(\rho)}_\tau\hspace*{0.2cm} \Phi^{(v)}\quad P^{(v)}_u\hspace*{0.2cm}
\Phi\hspace*{0.8cm}
\bar P^{(v)}_u\hspace*{0.4cm} \bar\Phi\hspace*{0.7cm} \Delta_I\quad  T^{v}_I\hspace*{0.8cm} \Delta\quad \Xi\hspace*{0.5cm}
\bar{\Delta}\quad \bar\Xi\hspace*{0.5cm} \Psi^{(v)}_{I}}\\ \cline{2-14}
P^{(\rho)}_{\tau} &0&-1&&&&&&&&&&&\\
\Phi^{(v)}&1&0&&&&&&&&&&&\\
P^{(v)}_u &&&0&2\rho^\tau&&&&&&&&&\\
\Phi&&&-2\rho^\tau&0&&&&&&&&&\\
\bar P^{(v)}_u &&&&&0&2\rho^\tau &&&&{\mathbf 0}&&&\\
\bar\Phi &&&&&-2\rho^\tau &0&&&&&&&\\
\Delta_I &&&&&&&0&2\rho^\tau&&&&&\\
T^{(v)}_I &&&&&&&-2\rho^\tau &0&&&&&\\
\Delta &&&&&&&&&0&2&&&\\
\Xi &&&&{\mathbf 0}&&&&&-2&0&&&\\
\bar{\Delta}&&&&&&&&&&&0&2&\\
\bar\Xi &&&&&&&&&&&-2&0&\\
\Psi^{v}_{I}&&&&&&&&&&&&&-16i\rho^\tau \\ \cline{2-14}
\end{array}
\end{equation}
multiplied by $\delta^{p}(\vec\sigma-\vec\sigma')$. Then we find the determinant of the matrix $\hat{\mathbf C}$
\begin{equation}\label{136}
det\hat{\mathbf C}=i(4\rho^\tau)^7 .
\end{equation}
The inverse matrix $\hat{\mathbf C^{-1}}$ is used to construct the Dirac brackets
\begin{equation}\label{137}
\begin{array}{rl}
\{f(\sigma),\;\; g(\sigma^\prime)\}_{D.B.}=&\{f(\sigma ),\; g(\sigma^\prime)\}_{P.B.} \\[0.2cm]
-&
\sum\int d^{p}\sigma''
\{f(\sigma ),\;F(\sigma'')\}_{P.B.}
( \hat{\mathbf C}^{-1})^{FG}(\sigma'')\{G(\sigma'')\,,
\; g(\sigma^\prime)\}_{P.B.},
\end{array}
\end{equation}
where $F$ and $G$ contain the second-class constraint set forming $\hat{\mathbf C}$.

The inverse Dirac matrix $\hat{\mathbf
C}^{-1}$ is given by
\begin{equation}\label{138}
\hat{\mathbf C}^{-1}=\begin{array} {l|ccccccccccccc|}
\multicolumn{14}{c}{\hspace*{0.9cm} P^{(\rho)}_\tau\hspace*{0.2cm} \Phi^{(v)}\quad P^{(v)}_u\hspace*{0.2cm}
\Phi\hspace*{0.8cm}
\bar P^{(v)}_u\hspace*{0.4cm} \bar\Phi\hspace*{0.4cm} \Delta_I\quad  T^{v}_I\hspace*{0.8cm} \Delta\quad \Xi\hspace*{0.5cm}
\bar{\Delta}\quad \bar\Xi\hspace*{0.5cm} \Psi^{(v)}_{I}}\\ \cline{2-14}
P^{(\rho)}_{\tau} &0&1&&&&&&&&&&&\\
\Phi^{(v)}&-1&0&&&&&&&&&&&\\
P^{(v)}_u &&&0&-\frac{1}{2\rho^\tau}&&&&&&&&&\\
\Phi&&&\frac{1}{2\rho^\tau}&0&&&&&&&&&\\
\bar P^{(v)}_u &&&&&0&-\frac{1}{2\rho^\tau} &&&&{\mathbf 0}&&&\\
\bar\Phi &&&&&\frac{1}{2\rho^\tau}&0&&&&&&&\\
\Delta_I &&&&&&&0&-\frac{1}{2\rho^\tau}&&&&&\\
T^{(v)}_I &&&&&&&\frac{1}{2\rho^\tau} &0&&&&&\\
\Delta &&&&&&&&&0&-\frac12&&&\\
\Xi &&&&{\mathbf 0}&&&&&\frac12&0&&&\\
\bar{\Delta}&&&&&&&&&&&0&-\frac12&\\
\bar\Xi &&&&&&&&&&&\frac12&0&\\
\Psi^{v}_{I}&&&&&&&&&&&&&\frac{i}{16\rho^\tau} \\ \cline{2-14}
\end{array}
\end{equation}\
multiplied by $\delta^{p}(\vec\sigma-\vec\sigma')$.

The matrices $\hat C$ (\ref{135}) and $\hat C^{-1}$ (\ref{138}) define the Hamiltonian symplectic structure in the total phase space of the original variables (\ref{20}) and (\ref{21}). It is easy to see that this phase space can be covariantly reduced by solving the constraint $\Phi^{(v)}$ which gives the following representation for the coordinate $\rho^\tau$
\begin{equation}\label{138'}
\rho^\tau=(vP\bar v).
\end{equation}
As a result, the canonical pair $(\rho^\tau, P^{(\rho)}_\tau)$ may be excluded from the original phase space,  because
$P^{(\rho)}_\tau\approx0$. This reduction does not change the matrix $\hat C$,  but only strikes out the upper $2\times2$ submatrix and substitutes $(vP\bar v)$ for $\rho^\tau$ in the remaining diagonal blocks. 
Then the partially reduced matrix $\hat C^{-1}_{red}$ is
substituted for $\hat C^{-1}$. As  a result of the  reduction the
matrix $\hat C^{-1}_{red}$ contains the nonlocal factor $\frac{1}{(vP\bar v)}$ absent in the D.B's. of the superparticle dynamics. This
peculiarity of the brane's D. B's. may lead to principal
distinctions between the string/brane and particle quantum
descriptions.

\section{Noncommutativity of coordinates under the Dirac brackets}

The D.B. (\ref{137}) defines the new commutation relations between the canonical variables encoding  the second-class constraint presence.

Analysis of the matrix  $\hat{\mathbf C}^{-1}$ (\ref{138}) structure shows that  the dyad-sector coordinates $u_\alpha, v_\beta$  commute among themselves
\begin{equation}\label{139}
\{u_\alpha(\vec\sigma), u_\beta(\vec\sigma')\}_{D.B.}=\{u_\alpha(\vec\sigma), v_\beta(\vec\sigma')\}_{D.B.}=\{v_\alpha(\vec\sigma), v_\beta(\vec\sigma')\}_{D.B.}=0
\end{equation}
and with $\rho^\tau$ and $\theta_\alpha$
\begin{equation}\label{140}
\begin{array}{c}
\{u_\alpha(\vec\sigma),\rho^\tau(\vec\sigma')\}_{D.B.}=0,\quad
\{v_\alpha(\vec\sigma),\rho^\tau(\vec\sigma')\}_{D.B.}=0, \\[0.2cm]
\{u_\alpha(\vec\sigma),\theta_\beta(\vec\sigma')\}_{D.B.}=0,\quad
\{v_\alpha(\vec\sigma),\theta_\beta(\vec\sigma')\}_{D.B.}=0.
\end{array}
\end{equation}
However, they do not commute  with $x_{\alpha\dot\alpha}$ and $z_{\alpha\beta}$
coordinates
\begin{equation}\label{141}
\begin{array}{c}
\{x_{\alpha\dot\alpha}(\vec\sigma), u_\beta(\vec\sigma')\}_{D.B.}=\frac{1}{\rho^\tau} u_\alpha \bar v_{\dot\alpha}v_\beta
\delta^p(\vec\sigma-\vec\sigma'), \nonumber\\[0.2cm]
\{x _{\alpha\dot\alpha}(\vec\sigma), v_\gamma(\vec\sigma')\}_{D.B.}=0,\nonumber\\[0.2cm]
\{ z_{\alpha\beta}(\vec\sigma), u_\gamma(\vec\sigma')\}_{D.B.}=\frac{1}{2\rho^\tau}v_\alpha v_\beta u_\gamma \delta^p(\vec\sigma-\vec\sigma'),
\nonumber\\[0.2cm]
\{z_{\alpha\beta}(\vec\sigma), v_\gamma(\vec\sigma')\}_{D.B.}=-
\frac{1}{2\rho^\tau}v_\alpha v_\beta v_\gamma\delta^p(\vec\sigma-\vec\sigma').
\end{array}
\end{equation}

The coordinates $\theta_\alpha$ have non-zero D.B's among themselves
\begin{eqnarray}\label{142}
\{\theta_{\alpha}(\vec\sigma),\theta_{\beta}(\vec\sigma')\}_{D.B.}={\textstyle\frac{i}{16\rho^\tau}}v_{\alpha} v_{\beta} \delta^p(\vec\sigma-\vec\sigma'),
\nonumber\\
\{\theta_\alpha(\vec\sigma),\bar\theta_{\dot\beta}(\vec\sigma')\}_{D.B.}=-{\textstyle\frac{i}{16\rho^\tau}}v_\alpha \bar v_ {\dot\beta} \delta^p(\vec\sigma-\vec\sigma'),
\nonumber\\
\{\bar\theta_{\dot\alpha}(\vec\sigma),\bar\theta_{\dot\beta}(\vec\sigma')\}_{D.B.}={\textstyle\frac{i}{16\rho^\tau}}\bar v_{\dot\alpha}\bar v_{\dot\beta} \delta^p(\vec\sigma-\vec\sigma')
\end{eqnarray}
and this results in the noncommutativity of the Goldstone fermion $\tilde\eta$
(\ref{3})
\begin{eqnarray}\label{143}
\{\theta_{\alpha}(\vec\sigma),\tilde\eta(\vec\sigma')\}_{D.B.}= {\textstyle\frac{1}{4\rho^\tau}}v_\alpha \delta^p(\vec\sigma-\vec\sigma'),\nonumber\\
\{\tilde\eta(\vec\sigma),\tilde\eta(\vec\sigma')\}_{D.B.}=-{\textstyle\frac{i}{\rho^\tau}} \delta^p(\vec\sigma-\vec\sigma').
\end{eqnarray}
But, the projection $(v^\alpha\theta_\alpha)$ associated with the unbroken directions  commutes with $\theta_\beta$ and  $\tilde\eta$
\begin{equation}\label{144}
\{v^\alpha\theta_\alpha(\vec\sigma),\theta_{\beta}(\vec\sigma')\}_{D.B.}=0,
\quad
\{v^\alpha\theta_\alpha(\vec\sigma),\tilde\eta(\vec\sigma')\}_{D.B.}=0
\end {equation}
The coordinates $\theta_\alpha$ have nonzero D.B's. with
$x_{\alpha\dot\alpha}$ and $z_{\alpha\beta}$ either:
\begin{eqnarray}\label{145}
\{\theta_{\alpha}(\vec\sigma),x_{\beta\dot\beta}(\vec\sigma')\}_{D.B.}={\textstyle\frac{1}{8\rho^\tau}}
v_\alpha (v_\beta \bar\theta_{\dot\beta}- \theta_{\beta}\bar v_{\dot\beta}) \delta^p(\vec\sigma-\vec\sigma'),\nonumber\\
\{\theta_{\alpha}(\vec\sigma), z_{\beta\delta}(\vec\sigma')\}_{D.B.}={\textstyle\frac{1}{8\rho^\tau}}
v_\alpha (\theta_{\beta} v_\delta + v_{\beta}\theta_{\delta}) \delta^p(\vec\sigma-\vec\sigma'),\\
\{\theta_{\alpha}(\vec\sigma), \bar z_{\dot\beta\dot\delta}(\vec\sigma')\}_{D.B.}=-{\textstyle\frac{1}{8\rho^\tau}}
v_\alpha (\bar\theta_{\dot\beta}\bar v_{\dot\delta} +\bar v_{\dot\beta}\bar\theta_{\dot\delta}) \delta^p(\vec\sigma-\vec\sigma').\nonumber
\end{eqnarray}
However, the projection $(v^\alpha\theta_\alpha)$ has  zero D.B's.  with the $x_{\alpha\dot\alpha}$ coordinates
\begin{equation}\label{146}
\{v^\alpha\theta_\alpha(\vec\sigma),x_{\beta\dot\beta}(\vec\sigma')\}_{D.B.}=0,
\end{equation}
but does not commute with the TCC coordinates  $z_{\beta\delta}$
\begin{equation}\label{147}
\{v^\alpha\theta_\alpha(\vec\sigma),z_{\beta\delta}(\vec\sigma')\}_{D.B.}={\textstyle\frac{1}{2\rho^\tau}}(v^\alpha \theta_{\alpha})v_\beta v_\delta \delta^p(\vec\sigma-\vec\sigma').
\end{equation}

The space-time $x_{\alpha\dot\alpha}$ and TCC coordinates $z_{\alpha\beta}$ have nonzero D.B'.s among themselves
\begin{equation}\label{148}
\begin{array}{c}
\{x_{\alpha\dot\alpha}(\vec\sigma),x_{\beta\dot\beta}(\vec\sigma')\}_{D.B.}=\frac{i}{4\rho^\tau}(\theta_\alpha\bar v_{\dot\alpha}-v_\alpha\bar\theta_{\dot\alpha})(\theta_\beta\bar v_{\dot\beta}-v_\beta\bar\theta_{\dot\beta})\delta^p(\vec\sigma-\vec\sigma'),
\\[0.2cm]
\{z_{\alpha\beta}(\vec\sigma),z_{\gamma\delta}(\vec\sigma')\}_{D.B.}=\frac{i}{4\rho^\tau}(\theta_\alpha v_{\beta}+\theta_\beta v_{\alpha})(\theta_\gamma v_{\delta}+\theta_\delta v_{\gamma})\delta^p(\vec\sigma-\vec\sigma'),
\\[0.2cm]
\{x_{\alpha\dot\alpha}(\vec\sigma),z_{\beta\gamma}(\vec\sigma')\}_{D.B.}=-\frac{i}{4\rho^\tau}(\theta_\alpha\bar v_{\dot\alpha}-v_\alpha\bar\theta_{\dot\alpha})(\theta_\beta v_\gamma+\theta_\gamma v_\beta)\delta^p(\vec\sigma-\vec\sigma').
\end{array}
\end{equation}

 The D.B. noncommutativity (\ref{148}) has gauge-dependent character and one can show that the r.h.s. of (\ref{148}) vanishes in the partially fixed $\kappa$-symmetry gauge
\begin{equation}\label{149}
\theta^{(v)}\equiv\theta^\alpha v_\alpha=0.
\end{equation}
To prove this we use the decomposition of $\theta_\alpha$ in the dyad basis
\begin{equation}
\theta_\alpha=\theta^{(v)}u_\alpha-\theta^{(u)}v_\alpha=\theta^{(v)}u_\alpha-(Re\theta^{(u)}-{\textstyle\frac{i}{4}}\tilde\eta)v_\alpha
\end{equation}
and present the multipliers entering the r.h.s. of (\ref{148}) as
\begin{equation}\label{151}
\begin{array}{c}
(\theta_\alpha\bar v_{\dot\alpha}-v_\alpha\bar\theta_{\dot\alpha})=\frac{i}{2}\tilde\eta v_\alpha\bar v _{\dot\alpha}+i(Im\theta^{(v)}m^{(+)}_{\alpha\dot\alpha}-Re\theta^{(v)}m^{(-)}_{\alpha\dot\alpha}),\\[0.2cm]
(\theta_\beta v_\gamma+\theta_\gamma v_\beta)=\frac{i}{2}\tilde\eta v_\beta v_\gamma-2Re\theta^{(v)}v_\beta v_\gamma+\theta^{(v)}(u_\beta v_\gamma+u_\gamma v_\beta).\\
\end{array}
\end{equation}
In the gauge (\ref{149}) the representations (\ref{151}) reduce to
\begin{equation}\label{152}
\begin{array}{c}
(\theta_\alpha\bar v_{\dot\alpha}-v_\alpha\bar\theta_{\dot\alpha})|_{\mbox{\footnotesize gauge (\ref{149})}}=\frac{i}{2}\tilde\eta v_\alpha\bar v _{\dot\alpha},\\[0.2cm]
(\theta_\beta v_\gamma+\theta_\gamma v_\beta)|_{\mbox{\footnotesize gauge (\ref{149})}}=\frac{i}{2}\tilde\eta v_\beta v_\gamma\\
\end{array}
\end{equation}
and the substitution of (\ref{152}) into (\ref{148}) yields
\begin{equation}
\begin{array}{c}
\{x_{\alpha\dot\alpha}(\vec\sigma),x_{\beta\dot\beta}(\vec\sigma')\}_{D.B.}|_{\mbox{\footnotesize gauge (\ref{149})}}=\{z_{\alpha\beta}(\vec\sigma),z_{\gamma\delta}(\vec\sigma')\}_{D.B.}|_{\mbox{\footnotesize gauge (\ref{149})}}=0,\\[0.2cm]
\{x_{\alpha\dot\alpha}(\vec\sigma),z_{\beta\gamma}(\vec\sigma')\}_{D.B.}|_{\mbox{\footnotesize gauge (\ref{149})}}=0,
\end{array}
\end{equation}
in view of the relation $\tilde\eta^2=0$. This proves the gauge-dependent character of the D.B.  noncommutativity of the coordinates  $x$ and $z$ between themselves.

In contrast to this picture, the D.B.  noncommutativity of $\theta_\alpha$ with $x_{\alpha\dot\alpha}$ and $z_{\alpha\beta}$ coordinates cannot be removed by gauge fixing since
\begin{equation}
\begin{array}{c}
\{\theta_\alpha(\vec\sigma),x_{\beta\dot\beta}(\vec\sigma')\}_{D.B.}|_{\mbox{\footnotesize gauge (\ref{149})}}=\frac{i\tilde\eta}{16\rho^\tau}v_\alpha v_\beta\bar v _{\dot\beta}\delta^p(\vec\sigma-\vec\sigma'),\\[0.2cm]
\{\theta_\alpha(\vec\sigma),z_{\beta\gamma}(\vec\sigma')\}_{D.B.}|_{\mbox{\footnotesize gauge (\ref{149})}}=\frac{i\tilde\eta}{16\rho^\tau}v_\alpha v_\beta v _{\gamma}\delta^p(\vec\sigma-\vec\sigma').\\
\end{array}
\end{equation}
The same conclusion takes place for the D.B.'s of the light-like projection $(ux\bar u)$ with the transverse coordinates $x^{(+)}$ and $x^{(-)}$
\begin{equation}\label{155}
\begin{array}{c}
\{(ux\bar u)(\vec\sigma),x^{(+)}(\vec\sigma')\}_{D.B.}=(\frac{1}{\rho^\tau}x^{(+)}+\frac{\tilde\eta}{4}Re\theta^{(v)})\delta^p(\vec\sigma-\vec\sigma'),\\[0.2cm]
\{(ux\bar u)(\vec\sigma),x^{(-)}(\vec\sigma')\}_{D.B.}=(\frac{1}{\rho^\tau}x^{(-)}-\frac{\tilde\eta}{4}Im\theta^{(v)})\delta^p(\vec\sigma-\vec\sigma'),
\end{array}
\end{equation}
whereas
\begin{equation}\label{156}
\{x^{(+)}(\vec\sigma),x^{(-)}(\vec\sigma')\}_{D.B.}={\textstyle\frac{1}{\rho^\tau}}Re\theta^{(v)}Im\theta^{(v)}\delta^p(\vec\sigma-\vec\sigma'),
\end{equation}
which vanishes in the gauge (\ref{149}).
The D.B's. (\ref{155}), (\ref{156}) are accompanied with zero D.B.
\begin{equation}
\begin{array}{c}
\{(vx\bar v)(\vec\sigma),x^{(\pm)}(\vec\sigma')\}_{D.B.}=0,\\[0.2cm]
\{x^{(+)}(\vec\sigma),x^{(+)}(\vec\sigma')\}_{D.B.}=\{x^{(-)}(\vec\sigma),x^{(-)}(\vec\sigma')\}_{D.B.}=0,\\[0.2cm]
\{(ux\bar u)(\vec\sigma),(ux\bar u)(\vec\sigma')\}_{D.B.}=\{(vx\bar v)(\vec\sigma),(vx\bar v)(\vec\sigma')\}_{D.B.}=0,\\[0.2cm]
\{(ux\bar u)(\vec\sigma),(vx\bar v)(\vec\sigma')\}_{D.B.}=0.
\end{array}
\end{equation}
Note also that the $x_{\alpha\dot\alpha}$ components have  nonzero D.B's. with $\rho^\tau(\tau,\vec\sigma)$
\begin{equation}
\{x_{\alpha\dot\alpha}(\vec\sigma),\rho^\tau(\vec\sigma')\}_{D.B.}=v_\alpha\bar v_{\dot\alpha}\delta^p(\vec\sigma-\vec\sigma'),
\end{equation}
in contrast to the TCC coordinates
\begin{equation}
\{z_{\alpha\beta}(\vec\sigma),\rho^\tau(\vec\sigma')\}_{D.B.}=0.
\end{equation}
It shows that the classical variable $\rho^\tau$ has to be changed by the operator projection $(v\hat P\bar v)$ after quantization in the reduced phase
space, as it  follows from the constraint (\ref{138'}).
As a result, in the quantum brane dynamics some  commutation relations between
brane coordinates will be proportional to the non-local factor $\frac{1}{(vP\bar v)}$ due to the substitution of (anti)commutators for the D.B's.
The  vector $v_{\alpha}\bar v_{\dot \alpha}$ fixing the projection direction
 is a  light-like vector, so the projection  $(vP\bar v)$ has physical sense analogous to  $p_-$ (or $p_+$) for the Green-Schwarz superstring \cite{GSW}.
The non-local factor proportional to  $\frac{1}{p_{-}}$ appears in the
two-point function of the energy-momentum tensor for the
Green-Schwarz superstring, where it is associated with the Weyl
anomaly presence. Note also that the presence of $\frac{1}{(vP\bar v)}$ in
the brane model correlates with the anomalous factor appearing in
the quantum algebra \cite{ILST} of the conformal symmetry of
tensionless string quantized in the light-cone gauge. This
correlation is not so evident, because our
consideration is free of any gauge fixing.

In  view of the above mentioned observations and $OSp(1,8)$
invariance of our  brane model, the D.B. and commutator
realizations of the $OSp(1,8)$ superalgebra deserve to be
carefully studied. The next section will be devoted to this goal.

\section{Dirac bracket realization of the $OSp(1|{\sf 8})$ superalgebra}

Having proved the  $OSp(1|8)$ symmetry of the super $p$-brane action (\ref{14})
 one can construct it P.B. and D.B.  realizations  and to observe the position of the non-local factor 
$\frac{1}{(vP\bar v)}$ in its structural functions.

The  generator densities $Q^\alpha$ and $Q^{\dot\alpha}$ of the $N=1$ global supersymmetry are given by
\begin{eqnarray}\label{159}
Q^{\alpha}(\tau,\vec\sigma)=\pi^\alpha +2i\bar\theta_{\dot\alpha} P^{\dot\alpha\alpha}+
4i\pi^{\alpha\beta}\theta_{\beta}, \nonumber  \\
\bar Q^
{\dot\alpha}(\tau,\vec\sigma)=\bar\pi^{\dot\alpha} +2iP^{\dot\alpha\alpha}\theta_{\alpha}+
4i\bar\pi^{\dot\alpha\dot\beta}\bar\theta_{\dot\beta}
\end{eqnarray}
and their P.B's. have the standard form
\begin{eqnarray}\label{160}
\{ Q^\alpha(\vec\sigma), \bar Q^{\dot\alpha}(\vec\sigma')\}_{P.B.}
=4iP^{\dot\alpha\alpha}\delta^p(\vec\sigma-\vec\sigma'), \nonumber  \\
\{ Q^{\alpha}(\vec\sigma), Q^{\beta}(\vec\sigma')\}_{P.B.}=8i \pi^{\alpha\beta}\delta^p(\vec\sigma-\vec\sigma').
\end{eqnarray}
The "square roots" $S_\gamma$ and $\bar S_{\dot\gamma}$ of the generalized conformal boost densities $K_{\gamma\dot\gamma}$, $K_{\gamma\lambda}$
\begin{eqnarray}\label{161}
\{ S_{\gamma}(\vec\sigma), \bar S_{\dot\rho}(\vec\sigma')\}_{P.B.}
=4iK_{\gamma\dot\rho}\delta^p(\vec\sigma-\vec\sigma'), \nonumber  \\
\{ S_\gamma(\vec\sigma), S_{\lambda}(\vec\sigma') \}_{P.B.}=4iK_{\gamma\lambda}\delta^p(\vec\sigma-\vec\sigma')
\end{eqnarray}
are  given by
\begin{equation}\label{162}
\begin{array}{rl}
S_{\gamma}(\tau,\vec\sigma)=z_{\gamma\delta}Q^\delta + x_{\gamma\dot\delta}\bar Q^{\dot\delta}-2i\theta_\gamma(\theta_\delta\pi^\delta+\bar\theta_{\dot\delta}\bar\pi^{\dot\delta})+4i(u^\delta\theta_\delta-\bar u^{\dot\delta}\bar\theta_{\dot\delta})P_{u\gamma},
\\
\bar S_{\dot\gamma}(\tau,\vec\sigma)=\bar z_{\dot\gamma\dot\delta}\bar Q^{\dot\delta}+x_{\delta\dot\gamma}Q^{\delta}-2i\bar\theta_{\dot\gamma}(\theta_\delta\pi^\delta+\bar\theta_{\dot\delta}\bar\pi^{\dot\delta})-4i(u^\delta\theta_\delta-\bar u^{\dot\delta}\bar\theta_{\dot\delta})\bar P_{u\dot\gamma}.
\end{array}
\end{equation}
The generalized conformal boost densities $K_{\gamma\lambda}$, $K_{\gamma\dot\gamma}$ are  respectively presented as
\begin{equation}\label{163}
\begin{array}{rl}
K_{\gamma\lambda}(\tau,\vec\sigma)=& 2z_{\gamma\beta}z_{\lambda\delta}\pi^{\beta\delta}+2x_{\gamma\dot\beta}x_{\lambda\dot\delta}\bar\pi^{\dot\beta\dot\delta}+z_{\gamma\beta}x_{\lambda\dot\delta}P^{\dot\delta\beta}+x_{\gamma\dot\beta}z_{\lambda\delta}
P^{\dot\beta\delta}\\[0.2cm]
+& \theta_\lambda(z_{\gamma\delta}\pi^{\delta}+x_{\gamma\dot\delta}\bar\pi^{\dot\delta})+\theta_{\gamma}(z_{\lambda\delta}\pi^{\delta}+x_{\lambda\dot\delta}\bar\pi^{\dot\delta})
\\[0.2cm]
+& (u^{\delta} z_{\delta\lambda}-\bar u^{\dot\delta} x_{\lambda\dot\delta})P_{
u\gamma}+(u^\delta z_{\delta\gamma}-\bar u^{\dot\delta}x_{\gamma\dot\delta})P_{u\lambda}\\[0.2cm]
-& 2i(u^\delta\theta_\delta-\bar u^{\dot\delta}\bar\theta_{\dot\delta})(\theta_{\lambda}P_{u\gamma}+\theta_\gamma P_{u\lambda}),\\
\end{array}
\end{equation}
\begin{equation}\label{164}
\begin{array}{rl}
K_{\gamma\dot\gamma}(\tau,\vec\sigma)=& z_{\gamma\delta}\bar z_{\dot\gamma\dot\delta}P^{\dot\delta\delta}+x_{\gamma\dot\delta}x_{\delta\dot\gamma}P^{\dot\delta\delta}+2(z_{\gamma\delta}x_{\lambda\dot\gamma}\pi^{\delta\lambda}+x_{\gamma\dot\delta}\bar z_{\dot\gamma\dot\lambda}\bar\pi^{\dot\delta\dot\lambda})\\[0.2cm]
+& \theta_\gamma(\bar z_{\dot\gamma\dot\delta}\bar\pi^{\dot\delta}+x_{\delta\dot\gamma}\pi^{\delta})+\bar\theta_{\dot\gamma}(z_{\gamma\delta}\pi^{\delta}+x_{\gamma\dot\delta}\bar\pi^{\dot\delta})\\[0.2cm]
+& (u^{\delta} x_{\delta\dot\gamma}-\bar u^{\dot\delta}\bar z_{\dot\delta\dot\gamma})P_{u\gamma}+(x_{\gamma\dot\delta}\bar u^{\dot\delta}-z_{\gamma\delta}u^\delta)\bar P_{u\dot\gamma}\\[0.2cm]
-& 2i(u^\delta\theta_\delta-\bar u^{\dot\delta}\bar\theta_{\dot\delta})(\bar\theta_{\dot\gamma}P_{u\gamma}-\theta_\gamma \bar P_{u\dot\gamma}).\\
\end{array}
\end{equation}
Using the matrix multiplication agreement one can present (\ref{163}) and
  (\ref{164}) in more compact form
\begin{equation}\label{165}
\begin{array}{rl}
K_{\gamma\lambda}(\tau,\vec\sigma)=& 2(z\pi z)_{\gamma\lambda}+2(x\bar\pi x)_{\gamma\lambda}
+ (xPz)_{\gamma\lambda} + (xPz)_{\lambda\gamma}\\[0.2cm]
-& [(z\pi)_\gamma\theta_\lambda + (z\pi)_{\lambda}\theta_{\gamma}]
-[(x\bar\pi)_{\gamma}\theta_{\lambda} + (x\bar\pi)_{\lambda}\theta_{\gamma}]\\[0.2cm]
+& [P_{u\gamma}(zu)_{\lambda}+ P_{u\lambda}(zu)_{\gamma}]
- [P_{u\gamma}(x\bar u)_{\lambda}+ P_{u\lambda}(x\bar u)_{\gamma}]\\[0.2cm]
-& 2i((u\theta) - (\bar u\bar\theta))
(\theta_{\gamma}P_{u\lambda} +\theta_\lambda P_{u\gamma}),
\end{array}
\end{equation}
and respectively
\begin{equation}\label{166}
\begin{array}{rl}
K_{\gamma\dot\gamma}(\tau,\vec\sigma)=&(\bar zPz)_{\dot\gamma\gamma}
+(xPx)_{\gamma\dot\gamma} + 2(z\pi x +x\bar\pi\bar z)_{\gamma\dot\gamma}\\[0.2cm]
+& \theta_\gamma(\pi x +\bar z \bar\pi)_{\dot\gamma}
+\bar\theta_{\dot\gamma}(z\pi + x\bar\pi)_{\gamma}\\[0.2cm]
+& P_{u\gamma}(ux- \bar u\bar z)_{\dot\gamma}
+ (x\bar u -zu)_{\gamma}\bar P_{u\dot\gamma}\\[0.2cm]
-& 2i((u\theta) - (\bar u\bar\theta))
(\bar\theta_{\dot\gamma}P_{u\gamma} -\theta_\gamma \bar P_{u\dot\gamma}).
\end{array}
\end{equation}

The remaining 16 generator densities of the $Sp(8)$ algebra $L_{\alpha\beta}$, $L_{\gamma\dot\rho}$ (together with their complex conjugate) are defined by the P.B's.
\begin{equation}\label{167}
\{ Q^{\gamma}(\vec\sigma), S_{\rho}(\vec\sigma')\}_{P.B.}=4i{L^\gamma}_{\rho}\delta^p(\vec\sigma-\vec\sigma'), \,\;
\{ Q^{\gamma}(\vec\sigma), \bar S_{\dot\rho}(\vec\sigma')\}_{P.B.}=4iL^{\gamma}{}_{\dot\rho}\delta^p(\vec\sigma-\vec\sigma'),
\end{equation}
 and have the form
\begin{eqnarray}\label{168}
\nonumber
L^{\alpha}{}_{\beta}(\tau,\vec\sigma)=P^{\dot\beta\alpha}x_{\beta\dot\beta}+
2\pi^{\alpha\gamma}z_{\gamma\beta} - \pi^\alpha\theta_\beta +u^\alpha P_{u\beta},\\
L^\alpha{}_{\dot\beta}(\tau,\vec\sigma)= 2\pi^{\alpha\gamma}x_{\gamma\dot\beta} +
P^{\dot\gamma\alpha}\bar z_{\dot\beta\dot\gamma}
- \pi^\alpha\bar\theta_{\dot\beta} - u^\alpha \bar P_{u\dot\beta}
\end{eqnarray}
 completed by their complex conjugate P.B's.
The derived P.B. realization  of the  $OSp(1|8)$ superalgebra together with the definition of the Dirac brackets (\ref{137}) are enough for construction of the D.B. realization of the $OSp(1|8)$ superalgebra.

Keeping in mind  the results \cite{ILST} one can wait the anomaly
presence in the commutator of the operator densities
$K_{\gamma\dot\gamma}$ and  $ L^{\alpha}{}_\beta$.
To this end let us in  the first place calculate the Dirac bracket of the generalized superconformal boost density $K_{\gamma\dot\gamma}$ with  $ L^{\alpha}{}_\beta$. 
Then we find
\begin{equation}\label{169}
\begin{array}{rl}
\{K_{\gamma\dot\gamma}(\vec\sigma),L^{\alpha}{}_\beta(\vec\sigma')\}_{D.B.}
=&-\delta^\alpha_\gamma K_{\beta\dot\gamma}\delta^p(\vec\sigma-\vec\sigma')
\\[0.2cm]
-&\int d^{p}\sigma''  \{K_{\gamma\dot\gamma}(\vec\sigma),\Phi(\vec\sigma'')\}_{P.B.}(\frac{1}{2\rho^\tau(\vec\sigma'')})\{P^{(v)}_u(\vec\sigma''), L^\alpha{}_\beta(\vec\sigma')\}_{P.B.}\\[0.2cm]
+&\int d^{p}\sigma''\{K_{\gamma\dot\gamma}(\vec\sigma),\bar P^{(v)}_u(\vec\sigma'')\}_{P.B.}(\frac{1}{2\rho^\tau(\vec\sigma'')})\{\bar\Phi(\vec\sigma''), L^\alpha{}_\beta(\vec\sigma')\}_{P.B.}\\[0.2cm]
+
&\int d^{p}\sigma''
\{K_{\gamma\dot\gamma}(\vec\sigma),\Delta_I(\vec\sigma'')\}_{P.B.}
(\frac{1}{2\rho^\tau(\vec\sigma'')})
\{T^{(v)}_I(\vec\sigma''),
L^\alpha{}_\beta(\vec\sigma')\}_{P.B.}
\\[0.2cm]
-&\int d^{p}\sigma''\{K_{\gamma\dot\gamma}(\vec\sigma),\Psi^{(v)}_I(\vec\sigma'')\}_{P.B.}(\frac{i}{16\rho^\tau(\vec\sigma'')})\{\Psi^{(v)}_I(\vec\sigma''), L^\alpha{}_\beta(\vec\sigma')\}_{P.B.}\\
\end{array}
\end{equation}
with the omitted  terms vanishing in the strong sense. The terms in the r.h.s. of the D.B. (\ref{169}) containing  the non-local factor  $\frac{1}{\rho^\tau}$ have a special structure in view of which each of them is proportional to some of the constraints of the model.
As a result, we find that the contribution of the factor $\frac{1}{\rho^\tau}$ will
 be vanishing on the constraint surface.

To prove this fact we note that the second term in (\ref{169}) includes the two multipliers
\begin{equation}\label{170}
\begin{array}{rl}
\{K_{\gamma\dot\gamma}(\vec\sigma),\Phi(\vec\sigma'')\}_{P.B.}=&2[\rho^\tau u_\gamma u^\beta
(x_{\beta\dot\gamma}-2i\theta_\beta\bar\theta_{\dot\gamma})-\rho^\tau u_\gamma \bar u^{\dot\beta}
(\bar z_{\dot\beta\dot\gamma}-2i\bar\theta_{\dot\beta}\bar\theta_{\dot\gamma})
\\[0.2cm]
-& u_\gamma \bar v_{\dot\delta}\bar L^{\dot\delta}{}_{\dot\gamma}-\bar v_{\dot\gamma}u_\delta L^\delta{}_\gamma]\delta^p(\vec\sigma-\vec\sigma''),
\\[0.2cm]
\{P^{(v)}_u(\vec\sigma''), L^\alpha{}_\beta(\vec\sigma')\}_{P.B.}=& v^\alpha P_{u\beta}\delta^p(\vec\sigma''-\vec\sigma'),
\\[0.2cm]
\end{array}
\end{equation}
where $\bar L^{\dot\delta}{}_{\dot\gamma}$ is the Lorentz generator complex conjugate to $L^\delta{}_\gamma$. One can see that the third multiplier is vanished, because of the primary constraint $ P_{u\beta}\approx 0$ presence.

The third term in (\ref{169}) also includes the  multipliers equal
\begin{equation}\label{171}
\begin{array}{rl}
\{K_{\gamma\dot\gamma}(\vec\sigma),\bar P^{(v)}_u(\vec\sigma'')\}_{P.B.}
=& [\bar v^{\dot\delta}(\bar z_{\dot\delta\dot\gamma}
-2i\bar\theta_{\dot\delta}\bar\theta_{\dot\gamma})P_{u\gamma}
-\bar v^{\dot\delta}(x_{\gamma\dot\delta}
-2i\bar\theta_{\dot\delta}\theta_{\gamma})\bar P_{u\dot\gamma}]\delta^p(\vec\sigma-\vec\sigma''),
\\[0.2cm]
\{\bar\Phi(\vec\sigma''), L^\alpha{}_\beta(\vec\sigma')\}_{P.B.}=& 2v_\beta\bar u_{\dot\lambda}P^{\dot\lambda\alpha}\delta^p(\vec\sigma''-\vec\sigma')
\end{array}
\end{equation}
and the former is proportional to the primary constraint $P_{u\gamma} \approx 0$ and it complex conjugate. The same story is repeated for the fourth term in
(\ref{169}) which multipliers are equal to
\begin{equation}\label{172}
\begin{array}{rl}
\{K_{\gamma\dot\gamma}(\vec\sigma),\Delta_I(\vec\sigma'')\}_{P.B.}=& 2i[\bar u^{\dot\delta}(\bar z_{\dot\delta\dot\gamma}-2i\bar\theta_{\dot\delta}\bar\theta_{\dot\gamma})P_{u\gamma}-u^{\delta}(z_{\delta\gamma}-2i\theta_{\delta}\theta_{\gamma})
\bar P_{u\dot\gamma}]\delta^p(\vec\sigma-\vec\sigma''),\\[0.2cm]
\{T^{(v)}_I(\vec\sigma''), L^\alpha{}_\beta(\vec\sigma')\}_{P.B.}=& 2iv_\beta v_\lambda\pi^{\lambda\alpha}\delta^p(\vec\sigma''-\vec\sigma');
\end{array}
\end{equation}
and the first of them is vanished, because of the constraints
$P_{u\gamma}\approx 0$ and $\bar P_{u\dot\gamma}\approx 0$ presence.
Finally, the latter term in (\ref{169}) including the multipliers
\begin{equation}\label{173}
\begin{array}{rl}
\{K_{\gamma\dot\gamma}(\vec\sigma),\Psi^{(v)}_I(\vec\sigma'')\}_{P.B.}=&\left[8(\theta_\gamma\bar P_{u\dot\gamma}-
\bar\theta_{\dot\gamma}P_{u\gamma})-iv_\gamma\left((\bar z_{\dot\gamma\dot\delta}+2i\bar\theta_{\dot\gamma}\bar\theta_{\dot\delta})\bar\Psi^{\dot\delta}+(x_{\delta\dot\gamma}+2i\bar\theta_{\dot\gamma}\theta_{\delta})
\Psi^\delta\right)\right.
\\[0.2cm]
+& \left. i\bar v_{\dot\gamma}\left((z_{\gamma\delta}
+2i\theta_{\gamma}\theta_{\delta})\Psi^{\delta}+(x_{\gamma\dot\delta}+2i\theta_\gamma\bar\theta_{\dot\delta})\bar\Psi^{\dot\delta}\right)\right]\delta^p(\vec\sigma-\vec\sigma''),
\\
[0.2cm]
\{\Psi^{(v)}_I(\vec\sigma''), L^\alpha{}_\beta(\vec\sigma')\}_{P.B.}=& iv_\beta\Psi^\alpha
\delta^p(\vec\sigma''-\vec\sigma')
\\[0.2cm]
\end{array}
\end{equation}
is also vanished, because of the constraints $\Psi^\alpha\approx 0$ and $P_{u\gamma} \approx 0$ presence there.

Conclusion is  that the D.B. of  $K_{\gamma\dot\gamma}$ and $ L^{\alpha}{}_\beta$ on the surface of the  primary constraints coincides  with the
 Poisson bracket  and is equal to
\begin{equation}\label{174}
\{K_{\gamma\dot\gamma}(\vec\sigma),L^{\alpha}{}_\beta(\vec\sigma') \}_{D.B.}|_{\mbox{\footnotesize constraint surface}}=-\delta^\alpha_\gamma K_{\beta\dot\gamma}\delta^p(\vec\sigma-\vec\sigma').
\end{equation}

Similar analysis can be done for the D.B's. of the remaining generators of the $OSp(1|8)$ superalgebra and, as a result, one can find that the D.B and P.B realizations of the superalgebra coincide on the primary constraint surface.

 In the quantum dynamics the commutator $[\hat K_{\gamma\dot\gamma},\hat L^{\alpha}{}_\beta]$ has to be substituted for the D.B. (\ref{174}) and all the coordinates and momenta should be presented by the correspondent operators. Moreover, the  generators of the $OSp(1|8)$ quantum superalgebra have to include sums of the ordered products of the coordinates and momenta.
One can choose, e.g. $\hat{\cal Q}\hat{\cal P}$-ordering, where  $\hat{\cal Q}$ is a condensed  notation for a chain of coordinate operators $\hat q$ and, respectively, $\hat {\cal P}$ for a chain of momentum operators $\hat p$.
Then, taking into account that any, but fixed $\hat p$-operator is not contained two or more times into the  $\hat{\cal P}$-chains, forming the operators $\hat K_{\gamma\dot\gamma}$ and $\hat L^{\alpha}{}_\beta$, we find
\begin{equation}\label{175}
[\hat K_{\gamma\dot\gamma}(\vec\sigma),\hat L^{\alpha}{}_{\beta}(\vec\sigma')]|_{\mbox{\footnotesize constraint surface}}=-\delta^\alpha_\gamma \hat K_{\beta\dot\gamma}\delta^p(\vec\sigma-\vec\sigma'),
\end{equation}
 because the operator ordering was not broken during the calculation of the commutator (\ref{175}). Thus, this commutation relation is anomalous free on the primary constraint surface.

The same $\hat{\cal Q}\hat{\cal P}$-ordering procedure was applied to construct the remaining  quantum generators of the $OSp(1|8)$ superalgebra and we
 observed  that all (anti)commutators of the superalgebra generators are anomalous free on the surface of some of the primary constraints.

However, this observation is not yet enough to exclude a
possibility of quantum anomaly because, the above discussed subset
of the primary constraints contains not only the  second-class
constraints  but, also the first-class constraints which equal
zero only in the weak  sense. So,  more careful investigation of
this problem needs to be done and it will be presented in another
place.

\section{Conclusion}

 The Hamiltonian structure of the simplest $D=4$ $N=1$ super $p$-brane model of which general solution describes the  BPS state with exotic fraction of supersymmetry equal to $3/4$ was studied here.
The covariant division of the brane constraints into the first and
second classes was found. As a result, the generators of the local
symmetries and the covariant realization of the Dirac matrix $\hat
C$ were constructed. The matrix $\hat C$ was diagonalized and
presented in the symplectic form parametrized by the component
$\rho^\tau(\tau,\vec\sigma)$ of the brane world-volume vector
density $\rho^\mu(\tau,\vec\sigma)$. The corresponding D.B.
encoding  the Hamiltonian symplectic structure of the constrained
super $p$-brane dynamics were constructed. The D.B. commutation
relations between the original $p$-brane coordinates in the
centrally extended superspace were calculated and their D.B.
noncommutativity was established. The D.B. noncommutativity in the
subspace of the space-time and TCC coordinates was shown to have a
gauge dependent character and
 can be removed by  complete gauge fixing of the exotic $\kappa$-symmetry.
On the contrary, the D.B. noncommutativity of the Grassmannian
$\theta$ coordinates  among them and with the space-time and TCC
coordinates was shown to be gauge independent. The  constructed
Dirac brackets revealed a deep dynamical role of the original
auxiliary spinor variables which manifests itself in their
noncommutativity with the space-time and TCC coordinates. The same
effect was established for the D.B's. of the space-time coordinates
with $\rho^\tau$. Exclusion of the canonical pair $(\rho^\tau,
P^{(\rho)}_\tau)$ from the total phase space was shown to result
in the appearance of the non-local factor $\frac{1}{(vP\bar v)}$ in
the Dirac matrix $\hat C^{-1}_{red}$ and, consequently, in the
Dirac brackets. This peculiarity of the Dirac brackets changed by
(anti)commutators may turn out to be principal in the quantum
picture of the string/brane dynamics and we started the
investigation of the nonlocality  problem and calculated the Dirac
bracket realization of the global $OSp(1|8)$ superalgebra.
 Moreover, the  $\hat{\cal Q}\hat{\cal P}$-ordering procedure for the superalgebra  generators was  applied and shown was that the potentially dangerous
terms in the superalgebra commutators vanish on the primary constraint surface. Due to the first-class constraints presence among these primary constraints
a room for the anomaly presence is not yet excluded and a little bit more further investigation has to be done. This is an objective of our paper under preparation \cite{UZ4},
where the BRST procedure for the quantization of the considered string/brane model is studied. Note also, that our analysis is naturally applied for the super $p$-brane models preserving $\frac{{\sf M}-1}{{\sf M}}$ fraction of $N=1$ supersymmetry in higher dimensions. In particular, this concerns the centrally extended $D=11$ superspace, where $31/32$ fraction of $N=1$ supersymmetry  for the
studied here model of tensionless super $p$-brane is preserved  \cite{Bandos}. Some additional details arising at the transition from $D=4$ to $D=11$ can be found in the recent paper \cite{BdAPV}, where superstring model preserving less number fractions of $D=11$ $N=1$ supersymmetry was considered.

\section{Acknowledgements}

A.Z. thanks Fysikum at the Stockholm University and Mittag-Leffler Institute for kind hospitality and I. Bengtsson, O. Laudal, U. Lindstrom and B. Sundborg  for useful discussions.
The work was supported in part by the Royal Swedish Academy of Sciences and SFFR of Ukraine under project 02.07/276.

\end{document}